\documentclass[aps,prl,superscriptaddress,longbibliography]{revtex4-1}

\usepackage{graphicx}
\usepackage{dcolumn}
\usepackage{bm}
\usepackage{amssymb}
\usepackage{microtype}
\usepackage{xfrac}
\usepackage{array}
\usepackage{gensymb}
\usepackage{xcolor}

\usepackage{xcolor}

\begin{document}

\title{Coexistence of metallic and nonmetallic properties in the pyrochlore Lu$_2$Rh$_2$O$_7$}

\author{Alannah~M.~Hallas}
\affiliation{Department of Physics and Astronomy, McMaster University, Hamilton L8S 4M1, Canada}
\affiliation{Department of Physics and Astronomy and Rice Center for Quantum Materials, Rice University, Houston, TX, 77005 USA}
\email[Email: ]{alannah.hallas@gmail.com}

\author{Arzoo~Z.~Sharma}
\affiliation{Department of Chemistry, University of Manitoba, Winnipeg R3T 2N2, Canada}

\author{Cole~Mauws}
\affiliation{Department of Chemistry, University of Manitoba, Winnipeg R3T 2N2, Canada}

\author{Qiang Chen}
\affiliation{Department of Physics and Astronomy, University of Tennessee-Knoxville, Knoxville 37996-1220, USA}

\author{Haidong~D.~Zhou}
\affiliation{Department of Physics and Astronomy, University of Tennessee-Knoxville, Knoxville 37996-1220, USA}

\author{Cui Ding}
\affiliation{Department of Physics, Zhejiang University, Hangzhou 310027, China}

\author{Zizhou Gong}
\affiliation{Department of Physics, Columbia University, New York, New York 10027, USA}

\author{Makoto~Tachibana}
\affiliation{National Institute for Materials Science, 1-1 Namiki, Tsukuba 305-0044, Ibaraki, Japan}

\author{Paul~M.~Sarte}
\affiliation{School of Chemistry, University of Edinburgh, Edinburgh EH9 3FJ, United Kingdom}
\affiliation{Centre for Science at Extreme Conditions, University of Edinburgh, Edinburgh EH9 3FD, United Kingdom}

\author{J.~Paul~Attfield}
\affiliation{School of Chemistry, University of Edinburgh, Edinburgh EH9 3FJ, United Kingdom}
\affiliation{Centre for Science at Extreme Conditions, University of Edinburgh, Edinburgh EH9 3FD, United Kingdom}

\author{Graeme~M.~Luke}
\affiliation{Department of Physics and Astronomy, McMaster University, Hamilton L8S 4M1, Canada}
\affiliation{TRIUMF, 4004 Wesbrook Mall, Vancouver, BC V6T 2A3, Canada}
\affiliation{Canadian Institute for Advanced Research, Toronto M5G 1Z7, Canada}

\author{Christopher~R.~Wiebe} 
\affiliation{Department of Physics and Astronomy, McMaster University, Hamilton L8S 4M1, Canada}
\affiliation{Department of Chemistry, University of Manitoba, Winnipeg R3T 2N2, Canada}
\affiliation{Canadian Institute for Advanced Research, Toronto M5G 1Z7, Canada}
\affiliation{Department of Chemistry, University of Winnipeg, Winnipeg R3B 2E9, Canada}
\email[Email: ]{chris.r.wiebe@gmail.com}

\date{\today}



\begin{abstract}
Transition metal oxides of the $4d$ and $5d$ block have recently become the targets of materials discovery, largely due to their strong spin-orbit coupling that can generate exotic magnetic and electronic states. Here we report the high pressure synthesis of Lu$_2$Rh$_2$O$_7$, a new cubic pyrochlore oxide based on $4d^5$ Rh$^{4+}$ and characterizations via thermodynamic, electrical transport, and muon spin relaxation measurements. Magnetic susceptibility measurements reveal a large temperature-independent Pauli paramagnetic contribution, while heat capacity shows an enhanced Sommerfeld coefficient, $\gamma$ = 21.8(1) mJ/mol-Rh K$^2$. Muon spin relaxation measurements confirm that Lu$_2$Rh$_2$O$_7$ remains paramagnetic down to 2~K. Taken in combination, these three measurements suggest that Lu$_2$Rh$_2$O$_7$ is a correlated paramagnetic metal with a Wilson ratio of $R_W = 2.5$. However, electric transport measurements present a striking contradiction as the resistivity of Lu$_2$Rh$_2$O$_7$ is observed to monotonically increase with decreasing temperature, indicative of a nonmetallic state. Furthermore, although the magnitude of the resistivity is that of a semiconductor, the temperature dependence does not obey any conventional form. Thus, we propose that Lu$_2$Rh$_2$O$_7$ may belong to the same novel class of non-Fermi liquids as the nonmetallic metal FeCrAs.
\end{abstract}
\maketitle

\section{Introduction}

The electronic and magnetic properties of many transition metal oxides can be understood by considering three energy scales: (i) the on-site Coulomb repulsion, Hubbard $U$, (ii) the band width, $W$ and (iii) the spin-orbit coupling, $\lambda$. The strong-coupling limit of this paradigm occurs for $3d$ oxides where Hubbard repulsion dominates ($U/W \gg 1$) and spin-orbit coupling ($\lambda \propto Z^4$) is negligible. As a consequence, Mott insulating states are observed in many materials that one would expect to be metals from band theory, as exemplified by the parent compounds of the high temperature cuprate superconductors~\cite{lee2006doping}. The opposing limit occurs for $5d$ oxides, where the $d$ electron orbitals are far more spatially delocalized such that $U$ decreases while $W$ increases ($U/W \approx 1$). At the same time, the large mass enhancement brings spin-orbit coupling to the forefront, giving rise to remarkable correlated states such as topological insulators, Weyl semi-metals, and spin-orbit assisted Mott insulators~\cite{witczak2014correlated}. Intermediate to these two limits is the $4d$ oxides, where $U$, $W$, and $\lambda$ can all be comparable in magnitude. 

The chemically versatile pyrochlore oxides, $A_2B_2$O$_7$, are an excellent family of materials to explore the competition between $U$, $W$, and $\lambda$. In this structure, the $A$ site is usually occupied by a rare earth cation, while the $B$ site can be occupied by transition metals from the $3d$, $4d$, or $5d$ block (or a post-transition metal). Consequently, a wide assortment of electronic behaviors are observed, ranging from band insulators to correlated metals. Pyrochlores are also of significant interest for their magnetic properties. The corner-sharing tetrahedral arrangement of the $A$ and $B$ cation sublattices is prone to intense geometric frustration, which has been extensively studied in the insulating families~\cite{gardner2010magnetic}. The metallic pyrochlore oxides provide an intriguing platform for studying the interplay of geometric frustration and metallic degrees of freedom, an emerging topic that has begun to garner significant attention~\cite{nakatsuji2006metallic,lacroix2010frustrated,julian2012frustrated,paddison2013emergent,billington2015magnetic,sapkota2017effective}. Whereas frustration in local moment systems can straightforwardly be identified via the so-called frustration index, $f = \sfrac{|\theta_{CW}|}{T_N}$~\cite{ramirez1994strongly}, no equivalent metric has yet been established for itinerant magnetic materials. Hence, how to unambiguously identify the presence of frustration in an itinerant system remains an open problem. A recent study broached this topic by using inelastic neutron scattering to extract exchange couplings which were then compared to a local moment picture~\cite{sapkota2017effective}.

In the rare earth pyrochlores, the $4f$ electrons from the $A$-site cation are highly localized and do not participate in electrical conduction. Thus, the electronic properties of pyrochlore materials are largely governed by the choice of $B$-site cation~\cite{subramanian1983oxide}. Insulating states are expected and observed in cases where there are no partially filled $d$-electron orbitals, such as when $B=$~Ti$^{4+}$, Ge$^{4+}$, Sn$^{4+}$, and Zr$^{4+}$. Due to their dominant Hubbard $U$, insulating or semiconducting states are also found in the majority of cases where $B$ has a partially occupied $3d$ shell, such as $B=$~Mn$^{4+}$ and Cr$^{4+}$. Semiconducting states are found for a number of pyrochlores with partially filled $4d$ or $5d$-shells, such as those with $B=$~Mo$^{4+}$, Ru$^{4+}$, and Pt$^{4+}$. Pyrochlores with metallic states are fewer in number, with notable examples being found amongst the $B=$~Os$^{5+}$, Re$^{5+}$, and Ir$^{4+}$ families. In these systems, large spin-orbit coupling results in fascinating correlated electron phenomena. For instance, the iridate family of pyrochlores have been lauded as candidates for topological electronic states~\cite{wan2011topological}. Meanwhile, Cd$_2$Re$_2$O$_7$ is a moderately heavy fermion metal, $\gamma$ = 13.3 mJ/mol-Re K$^2$~\cite{blacklock1979specific}, with a superconducting transition at $T_C = 1.47$~K~\cite{sakai2001superconductivity}. In the case of Cd$_2$Os$_2$O$_7$, the entanglement of spin, orbital, and lattice degrees of freedom drives a metal-to-insulator transition at 226~K concomitant with antiferromagnetic order~\cite{sleight1974semiconductor,mandrus2001continuous,calder2016spin,sohn2017strong}. Here, we introduce a new $4d$ pyrochlore, Lu$_2$Rh$_2$O$_7$, where we expect that $U$, $W$, and $\lambda$ are all comparable in magnitude, giving rise to an exotic non-Fermi liquid state.

The magnetism in Lu$_2$Rh$_2$O$_7$ originates from $4d^5$ Rh$^{4+}$ while $4f^{14}$ Lu$^{3+}$ is non-magnetic. Although several $4d^5$ rhodate pyrochlores have been previously reported~\cite{sleight1971platinum,lazarev1978electrical}, their characterization has been negligible, especially in contrast to their $5d^5$ iridate analogs. This can, in part, be attributed to the difficulty in stabilizing rhodium in its 4+ oxidation state, which we overcome using high pressure methods. In the pyrochlore lattice, Rh$^{4+}$ sits at the center of a trigonally distorted octahedral oxygen environment with $D_{3d}$ point group symmetry. As schematically presented in Figure~\ref{XRD}(a), the crystal electric field and moderately strong spin-orbit coupling ($\lambda = 0.19$~eV for Rh in contrast with $\lambda = 0.40$~eV for Ir~\cite{martins2017coulomb}) split the $d$ electron states into a filled $j_{\text{eff}} = 3/2$ band and a half-occupied $j_{\text{eff}} = 1/2$ band. Thus, akin to the $5d^5$ iridates~\cite{uematsu2015large}, the $4d^5$ rhodates can have a $j_{\text{eff}}=1/2$ magnetic degree of freedom, leading to a spin-orbit induced Mott insulating state~\cite{rau2016spin}. Both theoretical calculations and direct spectroscopic measurements have confirmed the applicability of the $j_{\text{eff}}=1/2$ picture in other rhodate materials~\cite{birol2015j,calder2016spin}. However, even in cases where $\lambda$ is not large enough to reach the pure $j_{\text{eff}}=1/2$ limit, as is likely the case in Lu$_2$Rh$_2$O$_7$, the influences of spin-orbit coupling are still strongly felt~\cite{martins2011reduced,luo2013li}. We have characterized the magnetic and electronic properties of Lu$_2$Rh$_2$O$_7$ with magnetic susceptibility, heat capacity, resistivity, and muon spin relaxation ($\mu$SR) measurements. These measurements reveal an apparent contradiction: magnetic susceptibility, heat capacity, and $\mu$SR are consistent with a strongly correlated paramagnetic metal. Resistivity data on the other hand suggests an unconventional semiconducting state and hence, a gap in the density of states. Thus, we propose that Lu$_2$Rh$_2$O$_7$ belongs to the class of ``nonmetallic metals'' a term coined to describe the behavior of FeCrAs~\cite{wu2009novel,akrap2014optical}.   

\section{Results}

\begin{figure}[]
\includegraphics[width=4.2in]{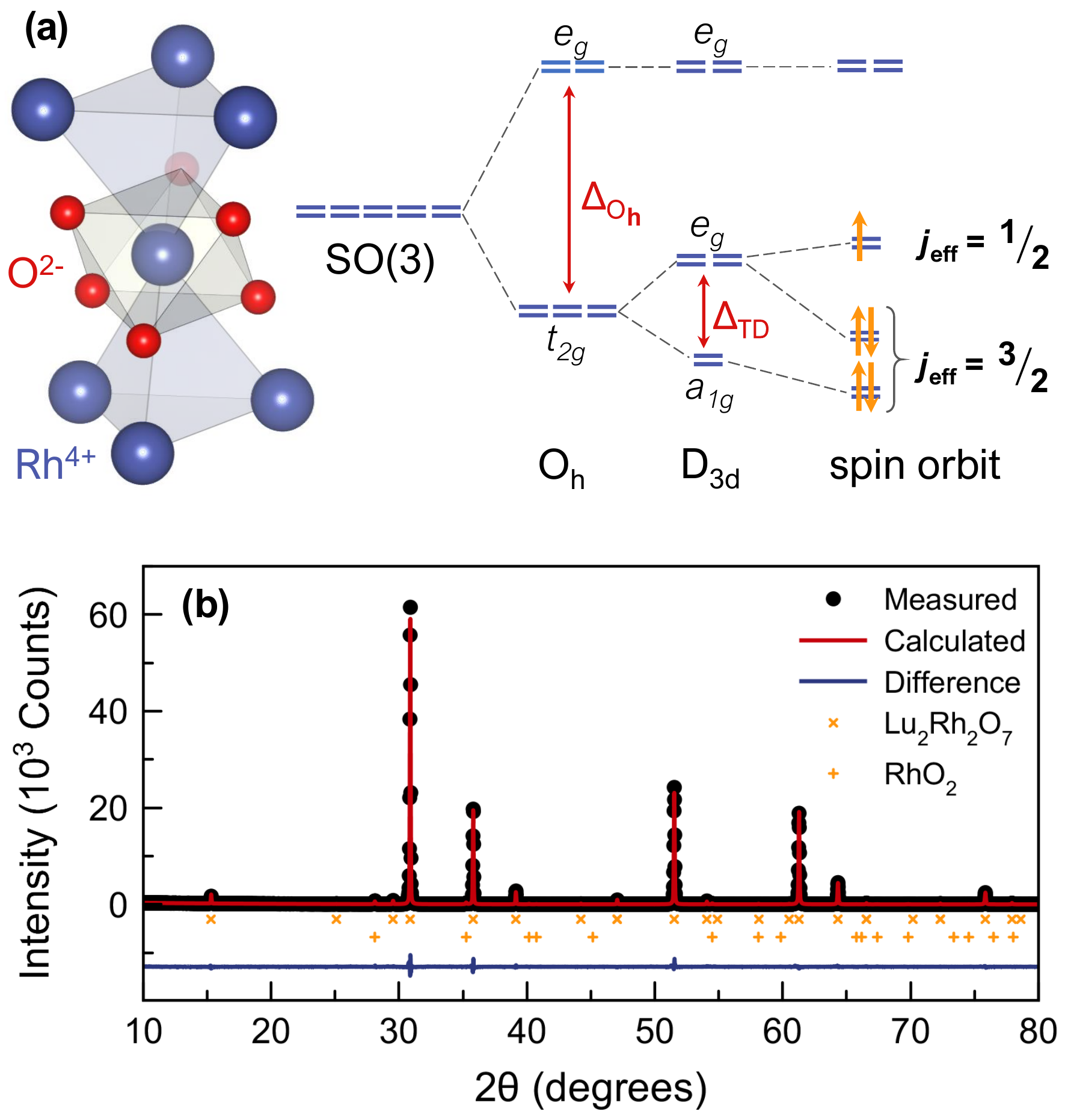}
\caption{(a) The $4d^5$ Rh$^{4+}$ cations in Lu$_2$Rh$_2$O$_7$ sit at the center of a trigonally distorted ($D_{3d}$) octahedral oxygen environment. Strong spin-orbit coupling will further split the $t_{2g}$ states into a filled $j_{\text{eff}}=3/2$ and a half-filled $j_{\text{eff}}=1/2$ state; however, this limit may not be fully reached for Rh$^{4+}$ with moderate spin-orbit coupling. (b) Powder x-ray diffraction pattern and Rietveld refinement for Lu$_2$Rh$_2$O$_7$ measured at $T = 300$~K with a copper $K_{\alpha1}$ wavelength ($\lambda = 1.5406$~\AA). The Bragg peak positions for the $Fd\bar{3}m$ pyrochlore phase are indicated by the upper set of crosses. The lower set of crosses are from a minor RhO$_2$ impurity accounting for 2\% of the sample volume. The results of the refinement are shown in Table I.}
\label{XRD}
\end{figure}

Under strongly oxidizing conditions, Lu$_2$Rh$_2$O$_7$ crystallizes in the pyrochlore structure (space group $Fd\bar{3}m$). The powder x-ray diffraction pattern, which is presented in Figure~\ref{XRD}(b), was analyzed via Rietveld refinement, giving a cubic lattice parameter of $a = 10.0265(1)$~\AA. The ratio of the ionic radii for Lu$^{3+}$ (86.1~pm) and Rh$^{4+}$ (60~pm) is within the stability range for the pyrochlore structure~\cite{subramanian1983oxide}. Thus, significant levels of $A/B$ site mixing are neither expected nor observed. The only adjustable coordinate within the pyrochlore structure is the O2 $x$ position, which we refine to $x = 0.3296(5)$. When $x = 0.3125$, the oxygen environment at the $B$-site is an ideal octahedron (point group $O_h$). Deviations from this value induce a compression along the local [111] axis, as shown in Figure~\ref{XRD}(a), such that the six equidistant oxygen ions have alternating bond angles of 93.5\degree~and 86.5\degree~with the central Rh$^{4+}$ cation (point group $D_{3d}$). A small RhO$_2$ impurity, comprising 2.1(2)\% of the total sample volume, was included as a second phase in the Rietveld refinements (a tetragonal structure with space group P4$_2$/mnm)~\cite{muller1968formation,shannon1968synthesis}. RhO$_2$ is a high pressure metallic oxide phase which is insoluble in aqua regia and could not be removed from our samples. The results of the Rietveld refinement, including the goodness-of-fit parameters, are listed in Table I.

\begin{figure}[]
\includegraphics[width=4.2in]{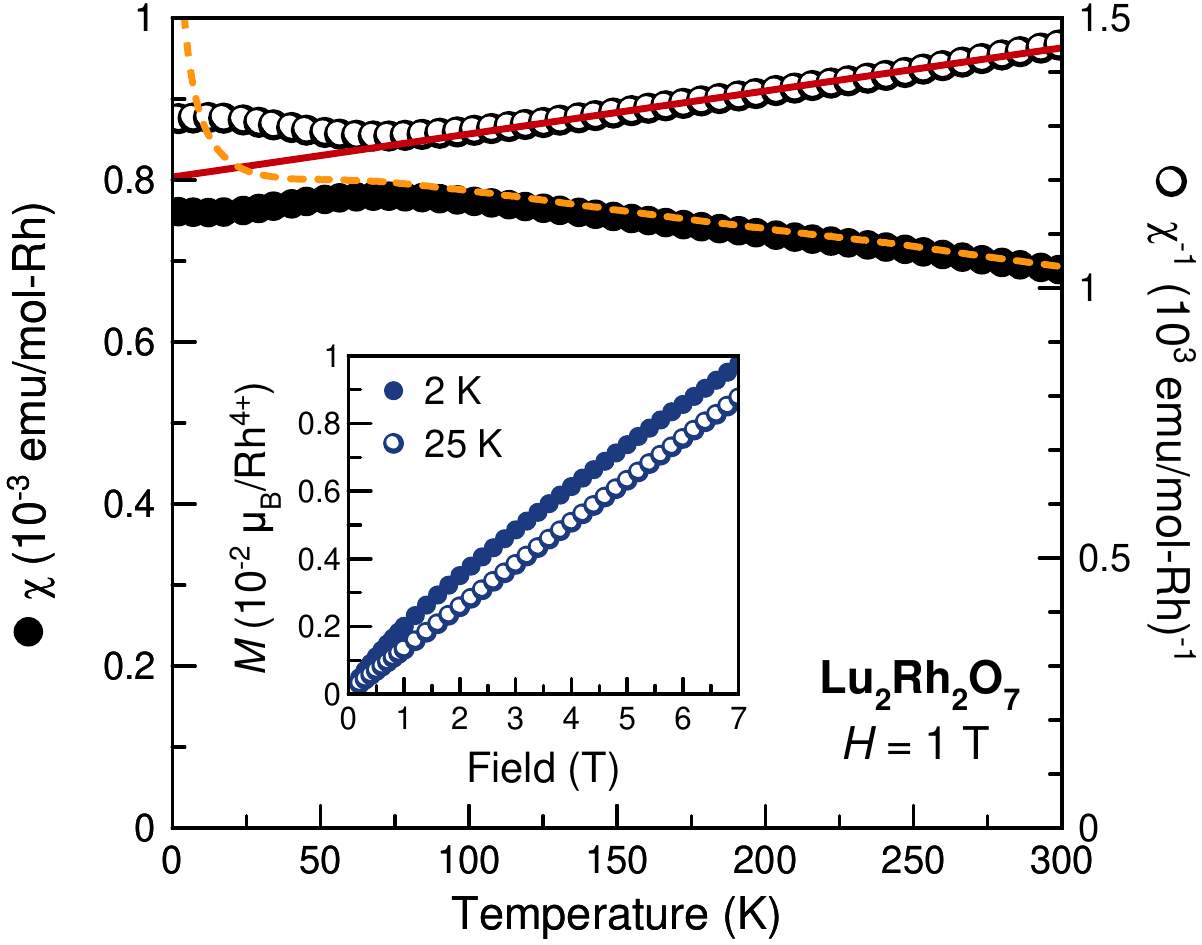}
\caption{The magnetic susceptibility of Lu$_2$Rh$_2$O$_7$, which was measured at $H = 1$~T, is dominated by a largely temperature independent Pauli paramagnetic contribution. The full symbols and the left hand axis are the susceptibility, $\chi$, while the open symbols and the right hand axis give the inverse susceptibility, $\chi^{-1}$. The yellow dashed line is the as-measured susceptibility data, prior to the Curie tail subtraction. The red line is a fit to the Curie-Weiss equation. The inset shows magnetization measurements at $T = 2$ and 25~K.}
\label{Susceptibility}
\end{figure}

The magnetic susceptibility of Lu$_2$Rh$_2$O$_7$ measured in an $H = 1$~T field is presented in Figure~\ref{Susceptibility}. We have corrected the data for core diamagnetism using the values tabulated in Ref.~\cite{bain2008diamagnetic}, which gives $\chi_D = -1.54\cdot10^{-4}$~emu/F.U. There is no appreciable difference between data collected under zero-field cooled or field cooled conditions. Lu$_2$Rh$_2$O$_7$ is paramagnetic down to the lowest measured temperature, with no evidence for magnetic ordering or spin freezing down to 2~K. The susceptibility is dominated by a large nearly temperature independent Pauli paramagnetic contribution. Below 10~K, the intrinsic susceptibility is obscured by a Curie-tail. We observe a shallow, broad hump in the susceptibility centered near 50~K. As can be seen in Figure~\ref{Susceptibility}, the inverse susceptibility between 100 and 300~K is reasonably well-described by a Curie-Weiss equation, $\chi = \frac{C}{T-\theta}$, where we obtain $\theta=-1393(2)$~K and $\mu_{\text{eff}}=\sqrt{8C}=3.1(1)$~$\mu_{B}$. Clearly this magnitude of $\theta$ is not physically meaningful as the susceptibility is dominated by the Pauli paramagnetic contribution. However, attempts to model the susceptibility to a modified Curie-Weiss law, $\chi = \frac{C}{T-\theta} + \chi_0$, failed to produce a unique solution. Instead, we proceed by subtracting the Curie-tail contribution, which we fit to $\chi \propto \frac{C}{T}$, allowing us to estimate the low temperature Pauli paramagnetic contribution as $\chi_P = 7.6\cdot10^{-4}$~emu/mol-Rh. This value is nearly an order of magnitude larger than in typical metals, signaling strong electron correlation effects. The induced magnetization at 2 K and 7 T is only 0.01~$\mu_B$/Rh$^{4+}$, consistent with an itinerant paramagnetic state (inset of Fig.~\ref{Susceptibility}). The susceptibility and magnetization data for Lu$_2$Rh$_2$O$_7$ bear a striking resemblance to that of SrRhO$_3$, another Rh$^{4+}$-based paramagnetic metal~\cite{yamaura2001enhanced}. The unusual properties of SrRhO$_3$ have been argued to emerge as the result of proximity to a quantum critical point~\cite{singh2003prospects}.

\begin{figure}[]
\includegraphics[width=4.2in]{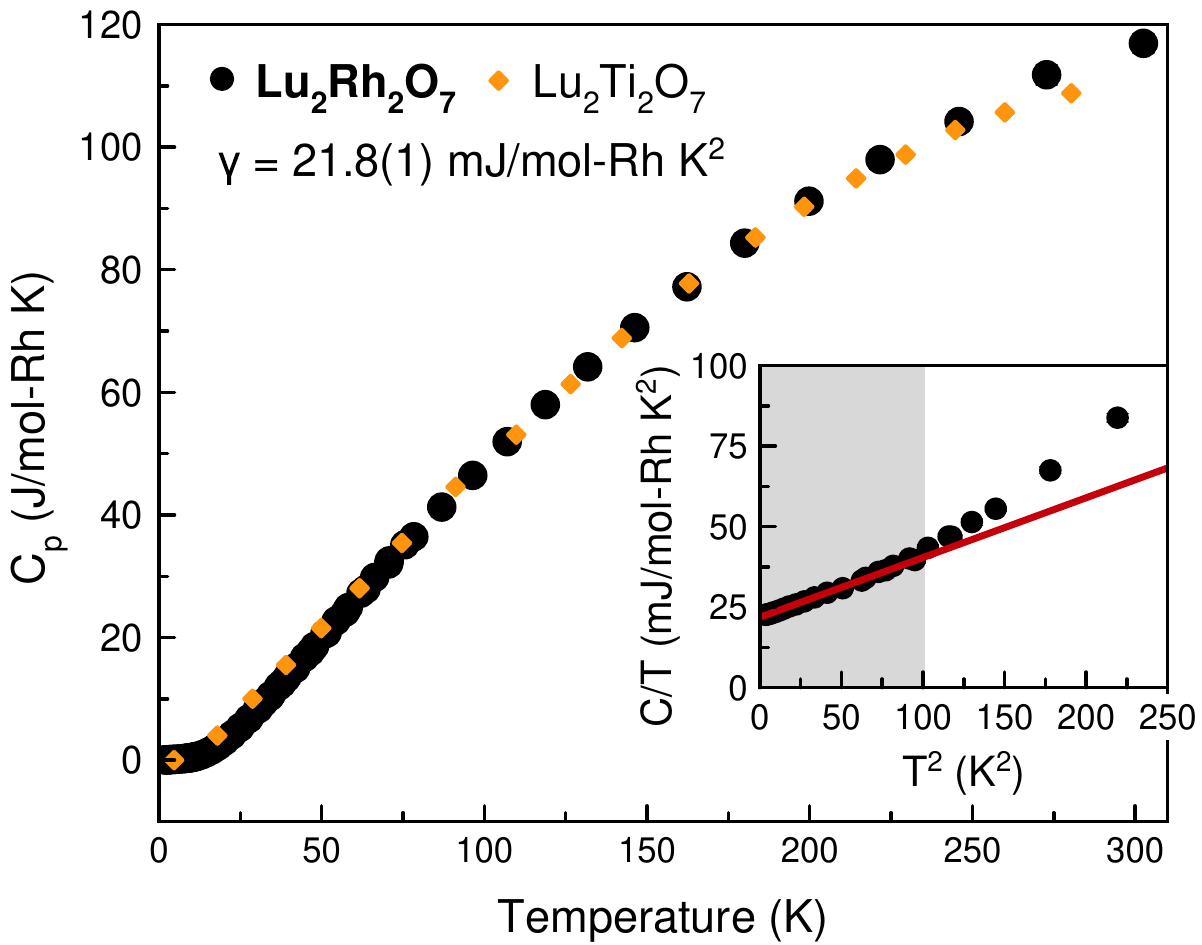}
\caption{The heat capacity of Lu$_2$Rh$_2$O$_7$ measured in zero field, with the scaled heat capacity of Lu$_2$Ti$_2$O$_7$ from Ref.~\cite{saha2008temperature} shown for reference. The inset shows a fit of $C/T = \gamma + \beta T^2$ over the shaded region, which yields a large Sommerfeld coefficient, $\gamma$ = 21.8(1) mJ/mol-Rh K$^2$.}
\label{heatcapacity}
\end{figure}

The heat capacity of Lu$_2$Rh$_2$O$_7$ is presented in Figure~\ref{heatcapacity}, where no phase transitions are observed between 2 and 300~K. For comparison, we plot alongside this data the scaled heat capacity of Lu$_2$Ti$_2$O$_7$, which is an insulating lattice analog with no unpaired $d$ electrons~\cite{saha2008temperature}. The heat capacity of Lu$_2$Ti$_2$O$_7$ was scaled by $\mu = \sqrt{\sfrac{M_{LRO}}{M_{LTO}}}$ where $M_{LRO}$ and $M_{LTO}$ are the molecular weights of Lu$_2$Rh$_2$O$_7$ and Lu$_2$Ti$_2$O$_7$, respectively. These two data sets are largely consistent, confirming that the high temperature heat capacity of Lu$_2$Rh$_2$O$_7$ can be accounted for by lattice degrees of freedom. The scaled heat capacity of Lu$_2$Ti$_2$O$_7$ is observed to slightly exceed that of Lu$_2$Rh$_2$O$_7$ below 50 K. This difference can be understood to originate from anharmonic effects, which are known to be significant in rare earth pyrochlores~\cite{maczka2008temperature}, and will vary between compounds dependent on the details of the phonon interactions. Indeed, the heat capacity of Lu$_2$Ti$_2$O$_7$ is also observed to exceed that of Dy$_2$Ti$_2$O$_7$ in the same temperature interval~\cite{saha2008temperature}. At the lowest temperatures, Fermi liquid behavior is observed for Lu$_2$Rh$_2$O$_7$, where $C/T$ is linear with respect to $T^2$ up to $T= 10$~K (Inset of Fig.~\ref{heatcapacity}). The fitted value for the Sommerfeld coefficient is $\gamma$ = 21.8(1) mJ/mol-Rh K$^2$, which is large and comparable to values observed in other strongly correlated rhodates and ruthenates~\cite{maeno1997two,park2016robust,perry2006sr2rho4}.

Enhanced electron masses are most commonly associated with heavy fermion physics in $f$-electron metals~\cite{stewart1984heavy}. However, there are several notable examples of $d$-electron systems with anomalously large Sommerfeld coefficients, including Y$_{1-x}$Sc$_x$Mn$_2$~\cite{shiga1993spin} and LiV$_2$O$_4$~\cite{kondo1997liv}. In both Y$_{1-x}$Sc$_x$Mn$_2$ (a cubic Laves phase) and LiV$_2$O$_4$ (a spinel), the underlying magnetic sublattice is a pyrochlore-like network of corner-sharing tetrahedra. It has been proposed that geometric frustration plays a central role in their electronic properties. Frustration tends to suppress conventional magnetic order, giving rise to strong spin fluctuations, while the itinerant nature of the electrons results in large spin disorder entropy at low temperatures, which gives rise to an enhanced linear contribution to the specific heat~\cite{lacroix2001heavy}. Indeed, the values for $\gamma$ in Y$_{1-x}$Sc$_x$Mn$_2$ (140 mJ/mol K$^2$ for $x=0.03$~\cite{wada1987spin}) and LiV$_2$O$_4$ (420 mJ/mol K$^2$~\cite{kondo1997liv}) are comparable to some of the best known examples of $f$-electron heavy fermion systems, such as URu$_2$Si$_2$ (180 mJ/mol K$^2$~\cite{palstra1985superconducting}) and UPt$_3$ (420 mJ/mol K$^2$~\cite{frings1983magnetic}). Given that the magnetic sublattice of Lu$_2$Rh$_2$O$_7$ shares this frustrated architecture, similar mechanisms may be responsible for its enhanced Sommerfeld coefficient.

\begin{figure}[]
\includegraphics[width=6.3in]{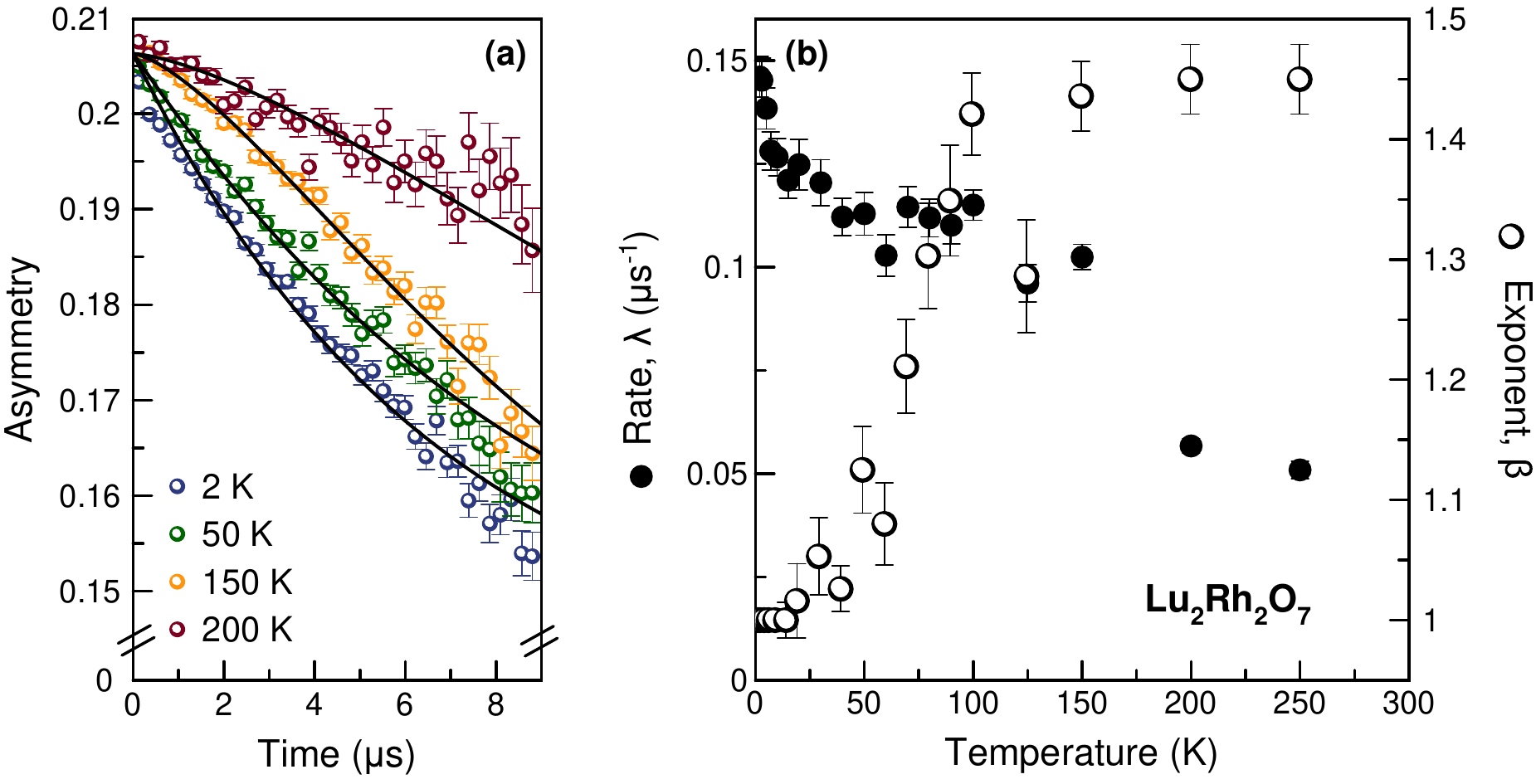} 
\caption{(a)Several representative asymmetry spectra from muon spin relaxation measurements on Lu$_2$Rh$_2$O$_7$. Error bars are derived from $\sqrt{N}$ counting statistics. The solid lines are fits to a stretched exponential function. (b) The temperature dependence of the fitted relaxation rate, $\lambda$, and the stretching parameter, $\beta$, revealing weak spin dynamics and a crossover from more Gaussian-like relaxation to Lorentzian relaxation.}
\label{muSR}
\end{figure}

We have also performed zero field $\mu$SR measurements on Lu$_2$Rh$_2$O$_7$. While the bulk probes employed thus far all indicate the absence of magnetic order, muons as a local probe are highly sensitive to very small ordered moments, even those below 0.1~$\mu_B$. Muon decay asymmetry spectra were collected between 2 and 250~K, and several representative data sets are presented in Figure~\ref{muSR}(a). The relaxation, which is weak at all measured temperatures, is observed to gradually increase with decreasing temperature. We do not observe any sharp increase in relaxation or spontaneous oscillations, as would be expected in the case of long-range magnetic order. As a function of temperature, the relaxation evolves from more Gaussian-like at high temperatures to Lorentzian at the lowest temperatures. Thus, we fitted the asymmetry spectra to a stretched exponential function
\begin{equation}
A(t) = A_0\cdot\exp{(-\lambda t)}^{\beta} + A_{BG}
\end{equation}
where $\lambda$ is the relaxation rate and the stretching parameter was constrained to lie between $\beta = 1$ (Lorentzian) and $\beta = 2$ (Gaussian). The non-relaxing background asymmetry, $A_{BG}$, accounts for muons which land outside the sample. The total asymmetry was constrained by fitting the precessing signal at high temperature in a weak field applied transverse to the muon's spin. The temperature dependence of both $\lambda$ and $\beta$ are presented in Figure~\ref{muSR}(b). The relaxation rate gradually grows below 200~K, reaching a maximum value of 0.15~$\mu s^{-1}$ at 2 K. There is a concomitant decrease of the stretching parameter, which plateaus at $\beta = 1$ below 20~K. This can be understood in terms of a crossover, where at high temperatures more Gaussian relaxation is observed due to fluctuating nuclear dipoles whereas at lower temperatures the signal is dominated by slowing spin dynamics. The decrease in $\beta$ may share an origin with the broad hump in the susceptibility in the same temperature interval. Moreover, these $\mu$SR measurements confirm that Lu$_2$Rh$_2$O$_7$ remains paramagnetic, with no evidence for magnetic order or spin freezing down to 2~K.


Electrical resistivity measurements were carried out using a conventional four-terminal method on a cold-pressed pellet~\cite{zhou2003transition} of Lu$_2$Rh$_2$O$_7$, which was annealed in air at 600~\degree C for 10 hours to minimize grain boundary effects. X-ray diffraction measurements were performed before and after the annealing protocol to confirm that it did not result in any degradation of the sample. In striking contrast to the magnetic susceptibility and heat capacity, which would both suggest a metallic state, the room temperature resistivity is 0.36 $\ohm$-cm, which is typical of a semiconducting state. Furthermore, the temperature dependence of the resistivity monotonically increases with decreasing temperature (\emph{ie.} $d\rho/dT < 0$). The resistivity is observed to diverge below 10~K. The inset of Figure~5 shows that this resistivity does not obey a thermally activated Arrhenius behavior for a very large range of temperatures. The data can be fit between 200 and 300~K to $\rho \propto \rho_a \cdot \exp(-\Delta / 2k_BT)$, yielding a very small gap of $\Delta = 37(1)$~meV. There are several factors worth considering in the quantitative interpretation of this data. First is that this sample is a pressed pellet and thus, scattering from grain boundaries may be expected to uniformly increase the resistivity. However, the cold-pressing and annealing protocol followed here has been demonstrated to provide nearly intrinsic transport measurements in other transition metal oxides, including samples that are far softer and more porous than the one measured here~\cite{zhou2003transition}. Moreover, even without the cold-pressing technique, conventional metallic resistivity (\emph{ie.} $d\rho/dT > 0$) has been measured in numerous polycrystalline samples of rhodates synthesized via the same high-pressure techniques used here~\cite{yamaura2001enhanced,yamaura2002crystal,yamaura2004crystal,yamaura2005high,yamaura2006high}. It is also worth mentioning that our minor impurity phase, RhO$_2$, is known to be a metal~\cite{shannon1968synthesis}. Thus, the nonmetallic resistivity is clearly intrinsic to Lu$_2$Rh$_2$O$_7$, a remarkable result in a material that all other probes suggest is metallic.

\section{Discussion}

Our characterizations have revealed that Lu$_2$Rh$_2$O$_7$ possesses a highly unusual non-Fermi liquid state. While both heat capacity and magnetic susceptibility indicate a finite density of states at the Fermi energy, the temperature dependence of the resistivity appears more consistent with a semiconducting gap. Moreover, even for a semiconductor, the temperature dependence of the resistivity is unconventional. It is instructive to consider a comparison with the ``nonmetallic metal'' FeCrAs. First, an important structural commonality is that in both materials, the magnetic constituent(s) are connected in triangular motifs that can be prone to geometric magnetic frustration. In the case of Lu$_2$Rh$_2$O$_7$ this is the corner-sharing tetrahedral pyrochlore sublattice and in the case of FeCrAs the non-magnetic Fe atoms occupy a triangular network of trimers while the magnetic Cr atoms form a distorted Kagome network. In regards to their electronic properties, the magnetic susceptibility of both materials is dominated by a largely temperature independent Pauli paramagnetic contribution~\cite{wu2009novel}. Secondly, both materials have a Sommerfeld coefficient, as determined by heat capacity measurements, significantly larger than typically expected for a $d$ electron system, $\gamma$ = 21.8(1) mJ/mol-Rh K$^2$ and 31.6 mJ/mol K$^2$ for Lu$_2$Rh$_2$O$_7$ and FeCrAs, respectively. However, the Wilson ratio ($R_W = 4\pi^2k_B^2\chi_P/3(g\mu_B)^2\gamma$) for FeCrAs, $R_W = 4-5$~\cite{wu2009novel} is larger than our computed value of $R_W = 2.5$ for Lu$_2$Rh$_2$O$_7$. A third commonality is the temperature dependence of the resistivity, where $d\rho/dT$ is negative, nominally indicative of a gap in the density of states. However, the room temperature resistivity for FeCrAs is approximately 0.4 m$\ohm$-cm, approximately three orders of magnitude smaller than our value of 0.4 $\ohm$-cm. Another significant difference between these two materials is that the Cr moments in FeCrAs undergo a long-range antiferromagnetic ordering transition at $T_N = 125$~K~\cite{wu2009novel}, while no magnetic order is observed in Lu$_2$Rh$_2$O$_7$ down to at least 2~K.

Several compelling explanations have been put forward to explain the nonmetallic metallic state in FeCrAs, which are worth examining in the current context of our results on Lu$_2$Rh$_2$O$_7$. The first proposal, which comes from Rau and Kee, suggests a hidden spin liquid state for the non-magnetic Fe trimers~\cite{rau2011hidden}. They propose that the electrons residing on these trimers are close to a metal-insulator quantum critical point, such that charge fluctuations strongly renormalize the transport behavior while only minimally affecting the thermodynamic properties. A second proposal that does not rely on proximity to a Mott insulating state, but instead attributes the unusual qualities of FeCrAs to it being a Hund's metal~\cite{nevidomskyy2009kondo}. In a Hund's metal, non-Fermi liquid properties emerge due to strong intra-atomic exchange in a multiorbital material, such as ruthenates and iron pnictides~\cite{georges2013strong}. The result of strong Hund's coupling in these materials is that a subset of the $d$ orbitals are pushed towards the local limit while maintaining strong coupling to other itinerant electrons. Several arguments make this second scenario less favorable in the case of Lu$_2$Rh$_2$O$_7$. First, Hund's coupling generally plays a weaker role in $4d^5$ Rh$^{4+}$~\cite{georges2013strong}. Secondly, this model is expected to give a large Wilson ratio~\cite{akrap2014optical}, as indeed observed in FeCrAs where $R_W = 4-5$~\cite{wu2009novel}, whereas in Lu$_2$Rh$_2$O$_7$, $R_W = 2.5$. Recent inelastic neutron scattering experiments have revealed high energy itinerant-like spin excitations, which Plumb \emph{et al.} propose may account for the nonmetallic resistivity~\cite{plumb2018mean}. Further study of the spin dynamics in Lu$_2$Rh$_2$O$_7$ are called for, a task which would be greatly benefited by the availability of large single crystal samples. While no single crystals of rhodate pyrochlores have been reported to date, a route similar to that used for Sr$_2$RhO$_4$ may prove fruitful~\cite{perry2006sr2rho4}.


Using high-pressure methods, we have synthesized a new pyrochlore Lu$_2$Rh$_2$O$_7$, in which $j_{\text{eff}}=1/2$ Rh$^{4+}$ occupies a frustrated corner-sharing tetrahedral lattice. No magnetic order is observed down to 2~K in $\mu$SR measurements. Heat capacity and magnetic susceptibility measurements both indicate strong electron correlations, with a large temperature independent Pauli paramagnetic response and an enhanced Sommerfeld coefficient, $\gamma$ = 21.8(1) mJ/mol-Rh K$^2$. Resistivity measurements, in contrast, suggest nonmetallic behavior ($d\rho/dT < 0$) in the full measured temperature range, 2 - 300~K. Thus, Lu$_2$Rh$_2$O$_7$ has a non-Fermi liquid state, sharing many commonalities with the nonmetallic metal FeCrAs.


\section{Methods}

High pressure synthetic techniques are often required to stabilize Rh$^{4+}$ compounds~\cite{muller1968formation,shannon1968synthesis,sleight1971platinum,lazarev1978electrical,yamaura2001enhanced,yamaura2002crystal,yamaura2004crystal}, as Rh$^{3+}$ is the more common oxidation state. The cubic pyrochlore Lu$_2$Rh$_2$O$_7$ was prepared using a belt-type high pressure apparatus at the National Institute for Materials Science in Tsukuba, Japan. Lu$_2$O$_3$ were thoroughly dried prior to the reaction. Rhodium metal was first oxidized to Rh$_2$O$_3$ by reacting under flowing oxygen at 1000~\degree C for 72 hours. The reaction into the pyrochlore phase was completed using KClO$_4$ as an oxidizing agent to promote the Rh$^{4+}$ oxidation state. The precursors (Lu$_2$O$_3$, Rh$_2$O$_3$, KClO$_4$ in a 1:1:$\frac{3}{8}$ ratio) were thoroughly ground together and sealed in a platinum capsule. The reaction was completed at 1700~\degree C and 6 GPa of applied pressure for 2 hours, after which the sample was quenched to room temperature before the pressure was released. The final product was repeatedly washed with water to remove KCl and then dried. The resulting 0.54~gram sample is made up of sub-micron sized black crystallites. Once prepared, this material is stable in air on a timescale of years. 

X-ray diffraction measurements were performed using a PANalytical diffractometer with copper K$_{\alpha 1}$ radiation, $\lambda = 1.540560$~\AA. Rietveld refinement was performed using FullProf~\cite{rodriguez1993recent}. Magnetic susceptibility, resistivity, and heat capacity measurements between 2 and 300~K were completed using a Quantum Design Physical Property Measurement System. Zero field $\mu$SR measurements were performed at the M20 surface muon channel at TRIUMF in the LAMPF Spectrometer. The $\mu$SR data were fitted over the full measured time interval, up to 9.5 $\mu$s, using the muSRfit software package.

\section{Data Availability}
The crystal structure of Lu$_2$Rh$_2$O$_7$, as determined by Rietveld refinement of the powder x-ray diffraction data, is available at the Cambridge Crystallographic Data Centre with deposition number CSD 1895045. All other data is available from the authors on reasonable request.

\section{Acknowledgments}

We gratefully acknowledge useful conversations with Yipeng Cai, Jonathan Gaudet, John Greedan, Chien-Lung Huang, and Murray Wilson. We thank Bassam Hitti and Gerald Morris from TRIUMF for their support with the $\mu$SR experiment. A.M.H. thanks the National Institute for Materials Science (NIMS) for their hospitality and support through the NIMS Internship Program. This work was supported by the Natural Sciences and Engineering Research Council of Canada (NSERC) and Canada Foundation for Innovation. C.R.W. is additionally supported by the Canada Research Chair (Tier II) program and the Canadian Institute for Advanced Research. Q.C. and H.D.Z. are supported by the National Science Foundation (NSF-DMR-1350002).  

\section{Competing Interests}

The authors declare no competing interests.

\section{Contributions}

The high-pressure synthesis and x-ray diffraction was performed by A.M.H. and M.T. Magnetic susceptibility and electric resistivity measurements, and their analysis, were performed by Q.C., H.D.Z., C.M., P.M.S., J.P.A., A.M.H., and C.R.W. Heat capacity measurements were performed by A.Z.S. and C.R.W. $\mu$SR measurements were performed and analyzed by A.M.H. and G.M.L. with assistance from C.D. and Z.G. The manuscript was written by A.M.H and C.R.W. with input from all authors.  

\bibliography{Rh_pyrochlores}

\begin{thebibliography}{55}%
\makeatletter
\providecommand \@ifxundefined [1]{%
 \@ifx{#1\undefined}
}%
\providecommand \@ifnum [1]{%
 \ifnum #1\expandafter \@firstoftwo
 \else \expandafter \@secondoftwo
 \fi
}%
\providecommand \@ifx [1]{%
 \ifx #1\expandafter \@firstoftwo
 \else \expandafter \@secondoftwo
 \fi
}%
\providecommand \natexlab [1]{#1}%
\providecommand \enquote  [1]{``#1''}%
\providecommand \bibnamefont  [1]{#1}%
\providecommand \bibfnamefont [1]{#1}%
\providecommand \citenamefont [1]{#1}%
\providecommand \href@noop [0]{\@secondoftwo}%
\providecommand \href [0]{\begingroup \@sanitize@url \@href}%
\providecommand \@href[1]{\@@startlink{#1}\@@href}%
\providecommand \@@href[1]{\endgroup#1\@@endlink}%
\providecommand \@sanitize@url [0]{\catcode `\\12\catcode `\$12\catcode
  `\&12\catcode `\#12\catcode `\^12\catcode `\_12\catcode `\%12\relax}%
\providecommand \@@startlink[1]{}%
\providecommand \@@endlink[0]{}%
\providecommand \url  [0]{\begingroup\@sanitize@url \@url }%
\providecommand \@url [1]{\endgroup\@href {#1}{\urlprefix }}%
\providecommand \urlprefix  [0]{URL }%
\providecommand \Eprint [0]{\href }%
\providecommand \doibase [0]{http://dx.doi.org/}%
\providecommand \selectlanguage [0]{\@gobble}%
\providecommand \bibinfo  [0]{\@secondoftwo}%
\providecommand \bibfield  [0]{\@secondoftwo}%
\providecommand \translation [1]{[#1]}%
\providecommand \BibitemOpen [0]{}%
\providecommand \bibitemStop [0]{}%
\providecommand \bibitemNoStop [0]{.\EOS\space}%
\providecommand \EOS [0]{\spacefactor3000\relax}%
\providecommand \BibitemShut  [1]{\csname bibitem#1\endcsname}%
\let\auto@bib@innerbib\@empty
\bibitem [{\citenamefont {Lee}\ \emph {et~al.}(2006)\citenamefont {Lee},
  \citenamefont {Nagaosa},\ and\ \citenamefont {Wen}}]{lee2006doping}%
  \BibitemOpen
  \bibfield  {author} {\bibinfo {author} {\bibfnamefont {PA}~\bibnamefont
  {Lee}}, \bibinfo {author} {\bibfnamefont {N}~\bibnamefont {Nagaosa}}, \ and\
  \bibinfo {author} {\bibfnamefont {X-G}\ \bibnamefont {Wen}},\ }\bibfield
  {title} {\enquote {\bibinfo {title} {Doping a {Mott} insulator: Physics of
  high-temperature superconductivity},}\ }\href@noop {} {\bibfield  {journal}
  {\bibinfo  {journal} {Reviews of Modern Physics}\ }\textbf {\bibinfo {volume}
  {78}},\ \bibinfo {pages} {17} (\bibinfo {year} {2006})}\BibitemShut {NoStop}%
\bibitem [{\citenamefont {Witczak-Krempa}\ \emph {et~al.}(2014)\citenamefont
  {Witczak-Krempa}, \citenamefont {Chen}, \citenamefont {Kim},\ and\
  \citenamefont {Balents}}]{witczak2014correlated}%
  \BibitemOpen
  \bibfield  {author} {\bibinfo {author} {\bibfnamefont {W}~\bibnamefont
  {Witczak-Krempa}}, \bibinfo {author} {\bibfnamefont {G}~\bibnamefont {Chen}},
  \bibinfo {author} {\bibfnamefont {YB}~\bibnamefont {Kim}}, \ and\ \bibinfo
  {author} {\bibfnamefont {L}~\bibnamefont {Balents}},\ }\bibfield  {title}
  {\enquote {\bibinfo {title} {Correlated quantum phenomena in the strong
  spin-orbit regime},}\ }\href@noop {} {\bibfield  {journal} {\bibinfo
  {journal} {Annu. Rev. Condens. Matter Phys.}\ }\textbf {\bibinfo {volume}
  {5}},\ \bibinfo {pages} {57--82} (\bibinfo {year} {2014})}\BibitemShut
  {NoStop}%
\bibitem [{\citenamefont {Gardner}\ \emph {et~al.}(2010)\citenamefont
  {Gardner}, \citenamefont {Gingras},\ and\ \citenamefont
  {Greedan}}]{gardner2010magnetic}%
  \BibitemOpen
  \bibfield  {author} {\bibinfo {author} {\bibfnamefont {JS}~\bibnamefont
  {Gardner}}, \bibinfo {author} {\bibfnamefont {MJP}\ \bibnamefont {Gingras}},
  \ and\ \bibinfo {author} {\bibfnamefont {JE}~\bibnamefont {Greedan}},\
  }\bibfield  {title} {\enquote {\bibinfo {title} {Magnetic pyrochlore
  oxides},}\ }\href@noop {} {\bibfield  {journal} {\bibinfo  {journal} {Reviews
  of Modern Physics}\ }\textbf {\bibinfo {volume} {82}},\ \bibinfo {pages} {53}
  (\bibinfo {year} {2010})}\BibitemShut {NoStop}%
\bibitem [{\citenamefont {Nakatsuji}\ \emph {et~al.}(2006)\citenamefont
  {Nakatsuji}, \citenamefont {Machida}, \citenamefont {Maeno}, \citenamefont
  {Tayama}, \citenamefont {Sakakibara}, \citenamefont {Van~Duijn},
  \citenamefont {Balicas}, \citenamefont {Millican}, \citenamefont {Macaluso},\
  and\ \citenamefont {Chan}}]{nakatsuji2006metallic}%
  \BibitemOpen
  \bibfield  {author} {\bibinfo {author} {\bibfnamefont {S}~\bibnamefont
  {Nakatsuji}}, \bibinfo {author} {\bibfnamefont {Y}~\bibnamefont {Machida}},
  \bibinfo {author} {\bibfnamefont {Y}~\bibnamefont {Maeno}}, \bibinfo {author}
  {\bibfnamefont {T}~\bibnamefont {Tayama}}, \bibinfo {author} {\bibfnamefont
  {T}~\bibnamefont {Sakakibara}}, \bibinfo {author} {\bibfnamefont
  {J}~\bibnamefont {Van~Duijn}}, \bibinfo {author} {\bibfnamefont
  {L}~\bibnamefont {Balicas}}, \bibinfo {author} {\bibfnamefont
  {JN}~\bibnamefont {Millican}}, \bibinfo {author} {\bibfnamefont
  {RT}~\bibnamefont {Macaluso}}, \ and\ \bibinfo {author} {\bibfnamefont
  {Julia~Y}\ \bibnamefont {Chan}},\ }\bibfield  {title} {\enquote {\bibinfo
  {title} {Metallic spin-liquid behavior of the geometrically frustrated kondo
  lattice {Pr$_2$Ir$_2$O$_7$}},}\ }\href@noop {} {\bibfield  {journal}
  {\bibinfo  {journal} {Physical Review Letters}\ }\textbf {\bibinfo {volume}
  {96}},\ \bibinfo {pages} {087204} (\bibinfo {year} {2006})}\BibitemShut
  {NoStop}%
\bibitem [{\citenamefont {Lacroix}(2010)}]{lacroix2010frustrated}%
  \BibitemOpen
  \bibfield  {author} {\bibinfo {author} {\bibfnamefont {Claudine}\
  \bibnamefont {Lacroix}},\ }\bibfield  {title} {\enquote {\bibinfo {title}
  {Frustrated metallic systems: A review of some peculiar behavior},}\
  }\href@noop {} {\bibfield  {journal} {\bibinfo  {journal} {Journal of the
  Physical Society of Japan}\ }\textbf {\bibinfo {volume} {79}},\ \bibinfo
  {pages} {011008} (\bibinfo {year} {2010})}\BibitemShut {NoStop}%
\bibitem [{\citenamefont {Julian}\ and\ \citenamefont
  {Kee}(2012)}]{julian2012frustrated}%
  \BibitemOpen
  \bibfield  {author} {\bibinfo {author} {\bibfnamefont {S}~\bibnamefont
  {Julian}}\ and\ \bibinfo {author} {\bibfnamefont {HY}~\bibnamefont {Kee}},\
  }\bibfield  {title} {\enquote {\bibinfo {title} {Frustrated metallic
  magnets},}\ }\href@noop {} {\bibfield  {journal} {\bibinfo  {journal}
  {Physics in Canada}\ }\textbf {\bibinfo {volume} {68}},\ \bibinfo {pages}
  {95} (\bibinfo {year} {2012})}\BibitemShut {NoStop}%
\bibitem [{\citenamefont {Paddison}\ \emph {et~al.}(2013)\citenamefont
  {Paddison}, \citenamefont {Stewart}, \citenamefont {Manuel}, \citenamefont
  {Courtois}, \citenamefont {McIntyre}, \citenamefont {Rainford},\ and\
  \citenamefont {Goodwin}}]{paddison2013emergent}%
  \BibitemOpen
  \bibfield  {author} {\bibinfo {author} {\bibfnamefont {Joseph~AM}\
  \bibnamefont {Paddison}}, \bibinfo {author} {\bibfnamefont {J~Ross}\
  \bibnamefont {Stewart}}, \bibinfo {author} {\bibfnamefont {Pascal}\
  \bibnamefont {Manuel}}, \bibinfo {author} {\bibfnamefont {Pierre}\
  \bibnamefont {Courtois}}, \bibinfo {author} {\bibfnamefont {Garry~J}\
  \bibnamefont {McIntyre}}, \bibinfo {author} {\bibfnamefont {Brian~D}\
  \bibnamefont {Rainford}}, \ and\ \bibinfo {author} {\bibfnamefont {Andrew~L}\
  \bibnamefont {Goodwin}},\ }\bibfield  {title} {\enquote {\bibinfo {title}
  {Emergent frustration in co-doped {$\beta$-Mn}},}\ }\href@noop {} {\bibfield
  {journal} {\bibinfo  {journal} {Physical Review Letters}\ }\textbf {\bibinfo
  {volume} {110}},\ \bibinfo {pages} {267207} (\bibinfo {year}
  {2013})}\BibitemShut {NoStop}%
\bibitem [{\citenamefont {Billington}\ \emph {et~al.}(2015)\citenamefont
  {Billington}, \citenamefont {Ernsting}, \citenamefont {Millichamp},
  \citenamefont {Lester}, \citenamefont {Dugdale}, \citenamefont {Kersh},
  \citenamefont {Duffy}, \citenamefont {Giblin}, \citenamefont {Taylor},
  \citenamefont {Manuel} \emph {et~al.}}]{billington2015magnetic}%
  \BibitemOpen
  \bibfield  {author} {\bibinfo {author} {\bibfnamefont {David}\ \bibnamefont
  {Billington}}, \bibinfo {author} {\bibfnamefont {David}\ \bibnamefont
  {Ernsting}}, \bibinfo {author} {\bibfnamefont {Thomas~E}\ \bibnamefont
  {Millichamp}}, \bibinfo {author} {\bibfnamefont {Christopher}\ \bibnamefont
  {Lester}}, \bibinfo {author} {\bibfnamefont {Stephen~B}\ \bibnamefont
  {Dugdale}}, \bibinfo {author} {\bibfnamefont {David}\ \bibnamefont {Kersh}},
  \bibinfo {author} {\bibfnamefont {Jonathan~A}\ \bibnamefont {Duffy}},
  \bibinfo {author} {\bibfnamefont {Sean~R}\ \bibnamefont {Giblin}}, \bibinfo
  {author} {\bibfnamefont {Jonathan~W}\ \bibnamefont {Taylor}}, \bibinfo
  {author} {\bibfnamefont {Pascal}\ \bibnamefont {Manuel}},  \emph {et~al.},\
  }\bibfield  {title} {\enquote {\bibinfo {title} {Magnetic frustration,
  short-range correlations and the role of the paramagnetic fermi surface of
  {PdCrO$_2$}},}\ }\href@noop {} {\bibfield  {journal} {\bibinfo  {journal}
  {Scientific reports}\ }\textbf {\bibinfo {volume} {5}},\ \bibinfo {pages}
  {12428} (\bibinfo {year} {2015})}\BibitemShut {NoStop}%
\bibitem [{\citenamefont {Sapkota}\ \emph {et~al.}(2017)\citenamefont
  {Sapkota}, \citenamefont {Ueland}, \citenamefont {Anand}, \citenamefont
  {Sangeetha}, \citenamefont {Abernathy}, \citenamefont {Stone}, \citenamefont
  {Niedziela}, \citenamefont {Johnston}, \citenamefont {Kreyssig},
  \citenamefont {Goldman} \emph {et~al.}}]{sapkota2017effective}%
  \BibitemOpen
  \bibfield  {author} {\bibinfo {author} {\bibfnamefont {A}~\bibnamefont
  {Sapkota}}, \bibinfo {author} {\bibfnamefont {BG}~\bibnamefont {Ueland}},
  \bibinfo {author} {\bibfnamefont {VK}~\bibnamefont {Anand}}, \bibinfo
  {author} {\bibfnamefont {NS}~\bibnamefont {Sangeetha}}, \bibinfo {author}
  {\bibfnamefont {DL}~\bibnamefont {Abernathy}}, \bibinfo {author}
  {\bibfnamefont {MB}~\bibnamefont {Stone}}, \bibinfo {author} {\bibfnamefont
  {JL}~\bibnamefont {Niedziela}}, \bibinfo {author} {\bibfnamefont
  {DC}~\bibnamefont {Johnston}}, \bibinfo {author} {\bibfnamefont
  {A}~\bibnamefont {Kreyssig}}, \bibinfo {author} {\bibfnamefont
  {AI}~\bibnamefont {Goldman}},  \emph {et~al.},\ }\bibfield  {title} {\enquote
  {\bibinfo {title} {Effective one-dimensional coupling in the highly
  frustrated square-lattice itinerant magnet {CaCo$_{2-y}$As$_2$}},}\
  }\href@noop {} {\bibfield  {journal} {\bibinfo  {journal} {Physical Review
  Letters}\ }\textbf {\bibinfo {volume} {119}},\ \bibinfo {pages} {147201}
  (\bibinfo {year} {2017})}\BibitemShut {NoStop}%
\bibitem [{\citenamefont {Ramirez}(1994)}]{ramirez1994strongly}%
  \BibitemOpen
  \bibfield  {author} {\bibinfo {author} {\bibfnamefont {AP}~\bibnamefont
  {Ramirez}},\ }\bibfield  {title} {\enquote {\bibinfo {title} {Strongly
  geometrically frustrated magnets},}\ }\href@noop {} {\bibfield  {journal}
  {\bibinfo  {journal} {Annual Review of Materials Science}\ }\textbf {\bibinfo
  {volume} {24}},\ \bibinfo {pages} {453--480} (\bibinfo {year}
  {1994})}\BibitemShut {NoStop}%
\bibitem [{\citenamefont {Subramanian}\ \emph {et~al.}(1983)\citenamefont
  {Subramanian}, \citenamefont {Aravamudan},\ and\ \citenamefont
  {Rao}}]{subramanian1983oxide}%
  \BibitemOpen
  \bibfield  {author} {\bibinfo {author} {\bibfnamefont {MA}~\bibnamefont
  {Subramanian}}, \bibinfo {author} {\bibfnamefont {G}~\bibnamefont
  {Aravamudan}}, \ and\ \bibinfo {author} {\bibfnamefont {GV~Subba}\
  \bibnamefont {Rao}},\ }\bibfield  {title} {\enquote {\bibinfo {title} {Oxide
  pyrochlores: A review},}\ }\href@noop {} {\bibfield  {journal} {\bibinfo
  {journal} {Progress in Solid State Chemistry}\ }\textbf {\bibinfo {volume}
  {15}},\ \bibinfo {pages} {55--143} (\bibinfo {year} {1983})}\BibitemShut
  {NoStop}%
\bibitem [{\citenamefont {Wan}\ \emph {et~al.}(2011)\citenamefont {Wan},
  \citenamefont {Turner}, \citenamefont {Vishwanath},\ and\ \citenamefont
  {Savrasov}}]{wan2011topological}%
  \BibitemOpen
  \bibfield  {author} {\bibinfo {author} {\bibfnamefont {X}~\bibnamefont
  {Wan}}, \bibinfo {author} {\bibfnamefont {AM}~\bibnamefont {Turner}},
  \bibinfo {author} {\bibfnamefont {A}~\bibnamefont {Vishwanath}}, \ and\
  \bibinfo {author} {\bibfnamefont {SY}~\bibnamefont {Savrasov}},\ }\bibfield
  {title} {\enquote {\bibinfo {title} {Topological semimetal and {Fermi}-arc
  surface states in the electronic structure of pyrochlore iridates},}\
  }\href@noop {} {\bibfield  {journal} {\bibinfo  {journal} {Physical Review
  B}\ }\textbf {\bibinfo {volume} {83}},\ \bibinfo {pages} {205101} (\bibinfo
  {year} {2011})}\BibitemShut {NoStop}%
\bibitem [{\citenamefont {Blacklock}\ and\ \citenamefont
  {White}(1979)}]{blacklock1979specific}%
  \BibitemOpen
  \bibfield  {author} {\bibinfo {author} {\bibfnamefont {K}~\bibnamefont
  {Blacklock}}\ and\ \bibinfo {author} {\bibfnamefont {HW}~\bibnamefont
  {White}},\ }\bibfield  {title} {\enquote {\bibinfo {title} {Specific heats of
  the pyrochlore compounds {Cd$_2$Re$_2$O$_7$} and {Cd$_2$Ru$_2$O$_7$}},}\
  }\href@noop {} {\bibfield  {journal} {\bibinfo  {journal} {The Journal of
  Chemical Physics}\ }\textbf {\bibinfo {volume} {71}},\ \bibinfo {pages}
  {5287--5289} (\bibinfo {year} {1979})}\BibitemShut {NoStop}%
\bibitem [{\citenamefont {Sakai}\ \emph {et~al.}(2001)\citenamefont {Sakai},
  \citenamefont {Yoshimura}, \citenamefont {Ohno}, \citenamefont {Kato},
  \citenamefont {Kambe}, \citenamefont {Walstedt}, \citenamefont {Matsuda},
  \citenamefont {Haga},\ and\ \citenamefont
  {Onuki}}]{sakai2001superconductivity}%
  \BibitemOpen
  \bibfield  {author} {\bibinfo {author} {\bibfnamefont {H}~\bibnamefont
  {Sakai}}, \bibinfo {author} {\bibfnamefont {K}~\bibnamefont {Yoshimura}},
  \bibinfo {author} {\bibfnamefont {H}~\bibnamefont {Ohno}}, \bibinfo {author}
  {\bibfnamefont {H}~\bibnamefont {Kato}}, \bibinfo {author} {\bibfnamefont
  {S}~\bibnamefont {Kambe}}, \bibinfo {author} {\bibfnamefont {RE}~\bibnamefont
  {Walstedt}}, \bibinfo {author} {\bibfnamefont {TD}~\bibnamefont {Matsuda}},
  \bibinfo {author} {\bibfnamefont {Y}~\bibnamefont {Haga}}, \ and\ \bibinfo
  {author} {\bibfnamefont {Y}~\bibnamefont {Onuki}},\ }\bibfield  {title}
  {\enquote {\bibinfo {title} {Superconductivity in a pyrochlore oxide,
  {Cd$_2$Re$_2$O$_7$}},}\ }\href@noop {} {\bibfield  {journal} {\bibinfo
  {journal} {Journal of Physics: Condensed Matter}\ }\textbf {\bibinfo {volume}
  {13}},\ \bibinfo {pages} {L785} (\bibinfo {year} {2001})}\BibitemShut
  {NoStop}%
\bibitem [{\citenamefont {Sleight}\ \emph {et~al.}(1974)\citenamefont
  {Sleight}, \citenamefont {Gillson}, \citenamefont {Weiher},\ and\
  \citenamefont {Bindloss}}]{sleight1974semiconductor}%
  \BibitemOpen
  \bibfield  {author} {\bibinfo {author} {\bibfnamefont {AW}~\bibnamefont
  {Sleight}}, \bibinfo {author} {\bibfnamefont {JL}~\bibnamefont {Gillson}},
  \bibinfo {author} {\bibfnamefont {JF}~\bibnamefont {Weiher}}, \ and\ \bibinfo
  {author} {\bibfnamefont {W}~\bibnamefont {Bindloss}},\ }\bibfield  {title}
  {\enquote {\bibinfo {title} {Semiconductor-metal transition in novel
  {Cd$_2$Os$_2$O$_7$}},}\ }\href@noop {} {\bibfield  {journal} {\bibinfo
  {journal} {Solid State Communications}\ }\textbf {\bibinfo {volume} {14}},\
  \bibinfo {pages} {357--359} (\bibinfo {year} {1974})}\BibitemShut {NoStop}%
\bibitem [{\citenamefont {Mandrus}\ \emph {et~al.}(2001)\citenamefont
  {Mandrus}, \citenamefont {Thompson}, \citenamefont {Gaal}, \citenamefont
  {Forro}, \citenamefont {Bryan}, \citenamefont {Chakoumakos}, \citenamefont
  {Woods}, \citenamefont {Sales}, \citenamefont {Fishman},\ and\ \citenamefont
  {Keppens}}]{mandrus2001continuous}%
  \BibitemOpen
  \bibfield  {author} {\bibinfo {author} {\bibfnamefont {D}~\bibnamefont
  {Mandrus}}, \bibinfo {author} {\bibfnamefont {JR}~\bibnamefont {Thompson}},
  \bibinfo {author} {\bibfnamefont {R}~\bibnamefont {Gaal}}, \bibinfo {author}
  {\bibfnamefont {L}~\bibnamefont {Forro}}, \bibinfo {author} {\bibfnamefont
  {JC}~\bibnamefont {Bryan}}, \bibinfo {author} {\bibfnamefont
  {BC}~\bibnamefont {Chakoumakos}}, \bibinfo {author} {\bibfnamefont
  {LM}~\bibnamefont {Woods}}, \bibinfo {author} {\bibfnamefont
  {BC}~\bibnamefont {Sales}}, \bibinfo {author} {\bibfnamefont
  {RS}~\bibnamefont {Fishman}}, \ and\ \bibinfo {author} {\bibfnamefont
  {V}~\bibnamefont {Keppens}},\ }\bibfield  {title} {\enquote {\bibinfo {title}
  {Continuous metal-insulator transition in the pyrochlore
  {Cd$_2$Os$_2$O$_7$}},}\ }\href@noop {} {\bibfield  {journal} {\bibinfo
  {journal} {Physical Review B}\ }\textbf {\bibinfo {volume} {63}},\ \bibinfo
  {pages} {195104} (\bibinfo {year} {2001})}\BibitemShut {NoStop}%
\bibitem [{\citenamefont {Calder}\ \emph {et~al.}(2016)\citenamefont {Calder},
  \citenamefont {Vale}, \citenamefont {Bogdanov}, \citenamefont {Liu},
  \citenamefont {Donnerer}, \citenamefont {Upton}, \citenamefont {Casa},
  \citenamefont {Said}, \citenamefont {Lumsden}, \citenamefont {Zhao} \emph
  {et~al.}}]{calder2016spin}%
  \BibitemOpen
  \bibfield  {author} {\bibinfo {author} {\bibfnamefont {S}~\bibnamefont
  {Calder}}, \bibinfo {author} {\bibfnamefont {JG}~\bibnamefont {Vale}},
  \bibinfo {author} {\bibfnamefont {NA}~\bibnamefont {Bogdanov}}, \bibinfo
  {author} {\bibfnamefont {X}~\bibnamefont {Liu}}, \bibinfo {author}
  {\bibfnamefont {C}~\bibnamefont {Donnerer}}, \bibinfo {author} {\bibfnamefont
  {MH}~\bibnamefont {Upton}}, \bibinfo {author} {\bibfnamefont {D}~\bibnamefont
  {Casa}}, \bibinfo {author} {\bibfnamefont {AH}~\bibnamefont {Said}}, \bibinfo
  {author} {\bibfnamefont {MD}~\bibnamefont {Lumsden}}, \bibinfo {author}
  {\bibfnamefont {Z}~\bibnamefont {Zhao}},  \emph {et~al.},\ }\bibfield
  {title} {\enquote {\bibinfo {title} {Spin-orbit-driven magnetic structure and
  excitation in the {$5d$} pyrochlore {Cd$_2$Os$_2$O$_7$}},}\ }\href@noop {}
  {\bibfield  {journal} {\bibinfo  {journal} {Nature Communications}\ }\textbf
  {\bibinfo {volume} {7}},\ \bibinfo {pages} {11651} (\bibinfo {year}
  {2016})}\BibitemShut {NoStop}%
\bibitem [{\citenamefont {Sohn}\ \emph {et~al.}(2017)\citenamefont {Sohn},
  \citenamefont {Kim}, \citenamefont {Sandilands}, \citenamefont {Hien},
  \citenamefont {Kim}, \citenamefont {Park}, \citenamefont {Kim}, \citenamefont
  {Moon}, \citenamefont {Yamaura}, \citenamefont {Hiroi} \emph
  {et~al.}}]{sohn2017strong}%
  \BibitemOpen
  \bibfield  {author} {\bibinfo {author} {\bibfnamefont {CH}~\bibnamefont
  {Sohn}}, \bibinfo {author} {\bibfnamefont {CH}~\bibnamefont {Kim}}, \bibinfo
  {author} {\bibfnamefont {LJ}~\bibnamefont {Sandilands}}, \bibinfo {author}
  {\bibfnamefont {NTM}\ \bibnamefont {Hien}}, \bibinfo {author} {\bibfnamefont
  {SY}~\bibnamefont {Kim}}, \bibinfo {author} {\bibfnamefont {HJ}~\bibnamefont
  {Park}}, \bibinfo {author} {\bibfnamefont {KW}~\bibnamefont {Kim}}, \bibinfo
  {author} {\bibfnamefont {SJ}~\bibnamefont {Moon}}, \bibinfo {author}
  {\bibfnamefont {J}~\bibnamefont {Yamaura}}, \bibinfo {author} {\bibfnamefont
  {Z}~\bibnamefont {Hiroi}},  \emph {et~al.},\ }\bibfield  {title} {\enquote
  {\bibinfo {title} {Strong spin-phonon coupling mediated by single ion
  anisotropy in the all-in--all-out pyrochlore magnet {Cd$_2$Os$_2$O$_7$}},}\
  }\href@noop {} {\bibfield  {journal} {\bibinfo  {journal} {Physical Review
  Letters}\ }\textbf {\bibinfo {volume} {118}},\ \bibinfo {pages} {117201}
  (\bibinfo {year} {2017})}\BibitemShut {NoStop}%
\bibitem [{\citenamefont {Sleight}\ and\ \citenamefont
  {Gillson}(1971)}]{sleight1971platinum}%
  \BibitemOpen
  \bibfield  {author} {\bibinfo {author} {\bibfnamefont {AW}~\bibnamefont
  {Sleight}}\ and\ \bibinfo {author} {\bibfnamefont {JL}~\bibnamefont
  {Gillson}},\ }\bibfield  {title} {\enquote {\bibinfo {title} {Platinum metal
  pyrochlores of the type {Tl$_2M_2$O$_7$}},}\ }\href@noop {} {\bibfield
  {journal} {\bibinfo  {journal} {Materials Research Bulletin}\ }\textbf
  {\bibinfo {volume} {6}},\ \bibinfo {pages} {781--784} (\bibinfo {year}
  {1971})}\BibitemShut {NoStop}%
\bibitem [{\citenamefont {Lazarev}\ and\ \citenamefont
  {Shaplygin}(1978)}]{lazarev1978electrical}%
  \BibitemOpen
  \bibfield  {author} {\bibinfo {author} {\bibfnamefont {VB}~\bibnamefont
  {Lazarev}}\ and\ \bibinfo {author} {\bibfnamefont {IS}~\bibnamefont
  {Shaplygin}},\ }\bibfield  {title} {\enquote {\bibinfo {title} {Electrical
  properties of mixed oxides containing a platinum metal and a non-noble
  metal},}\ }\href@noop {} {\bibfield  {journal} {\bibinfo  {journal} {Russ. J.
  Inorg. Chem}\ }\textbf {\bibinfo {volume} {23}},\ \bibinfo {pages} {163--170}
  (\bibinfo {year} {1978})}\BibitemShut {NoStop}%
\bibitem [{\citenamefont {Martins}\ \emph {et~al.}(2017)\citenamefont
  {Martins}, \citenamefont {Aichhorn},\ and\ \citenamefont
  {Biermann}}]{martins2017coulomb}%
  \BibitemOpen
  \bibfield  {author} {\bibinfo {author} {\bibfnamefont {C}~\bibnamefont
  {Martins}}, \bibinfo {author} {\bibfnamefont {M}~\bibnamefont {Aichhorn}}, \
  and\ \bibinfo {author} {\bibfnamefont {S}~\bibnamefont {Biermann}},\
  }\bibfield  {title} {\enquote {\bibinfo {title} {Coulomb correlations in
  {$4d$} and {$5d$} oxides from first principles—or how spin--orbit materials
  choose their effective orbital degeneracies},}\ }\href@noop {} {\bibfield
  {journal} {\bibinfo  {journal} {Journal of Physics: Condensed Matter}\
  }\textbf {\bibinfo {volume} {29}},\ \bibinfo {pages} {263001} (\bibinfo
  {year} {2017})}\BibitemShut {NoStop}%
\bibitem [{\citenamefont {Uematsu}\ \emph {et~al.}(2015)\citenamefont
  {Uematsu}, \citenamefont {Sagayama}, \citenamefont {Arima}, \citenamefont
  {Ishikawa}, \citenamefont {Nakatsuji}, \citenamefont {Takagi}, \citenamefont
  {Yoshida}, \citenamefont {Mizuki},\ and\ \citenamefont
  {Ishii}}]{uematsu2015large}%
  \BibitemOpen
  \bibfield  {author} {\bibinfo {author} {\bibfnamefont {D}~\bibnamefont
  {Uematsu}}, \bibinfo {author} {\bibfnamefont {H}~\bibnamefont {Sagayama}},
  \bibinfo {author} {\bibfnamefont {T}~\bibnamefont {Arima}}, \bibinfo {author}
  {\bibfnamefont {JJ}~\bibnamefont {Ishikawa}}, \bibinfo {author}
  {\bibfnamefont {S}~\bibnamefont {Nakatsuji}}, \bibinfo {author}
  {\bibfnamefont {H}~\bibnamefont {Takagi}}, \bibinfo {author} {\bibfnamefont
  {M}~\bibnamefont {Yoshida}}, \bibinfo {author} {\bibfnamefont
  {J}~\bibnamefont {Mizuki}}, \ and\ \bibinfo {author} {\bibfnamefont
  {K}~\bibnamefont {Ishii}},\ }\bibfield  {title} {\enquote {\bibinfo {title}
  {Large trigonal-field effect on spin-orbit coupled states in a pyrochlore
  iridate},}\ }\href@noop {} {\bibfield  {journal} {\bibinfo  {journal}
  {Physical Review B}\ }\textbf {\bibinfo {volume} {92}},\ \bibinfo {pages}
  {094405} (\bibinfo {year} {2015})}\BibitemShut {NoStop}%
\bibitem [{\citenamefont {Rau}\ \emph {et~al.}(2016)\citenamefont {Rau},
  \citenamefont {Lee},\ and\ \citenamefont {Kee}}]{rau2016spin}%
  \BibitemOpen
  \bibfield  {author} {\bibinfo {author} {\bibfnamefont {JG}~\bibnamefont
  {Rau}}, \bibinfo {author} {\bibfnamefont {EKH}\ \bibnamefont {Lee}}, \ and\
  \bibinfo {author} {\bibfnamefont {H-Y}\ \bibnamefont {Kee}},\ }\bibfield
  {title} {\enquote {\bibinfo {title} {Spin-orbit physics giving rise to novel
  phases in correlated systems: Iridates and related materials},}\ }\href@noop
  {} {\bibfield  {journal} {\bibinfo  {journal} {Annual Review of Condensed
  Matter Physics}\ }\textbf {\bibinfo {volume} {7}},\ \bibinfo {pages}
  {195--221} (\bibinfo {year} {2016})}\BibitemShut {NoStop}%
\bibitem [{\citenamefont {Birol}\ and\ \citenamefont
  {Haule}(2015)}]{birol2015j}%
  \BibitemOpen
  \bibfield  {author} {\bibinfo {author} {\bibfnamefont {T}~\bibnamefont
  {Birol}}\ and\ \bibinfo {author} {\bibfnamefont {K}~\bibnamefont {Haule}},\
  }\bibfield  {title} {\enquote {\bibinfo {title} {{$J_{\text{eff}} = 1/2$}
  {Mott}-insulating state in {Rh} and {Ir} fluorides},}\ }\href@noop {}
  {\bibfield  {journal} {\bibinfo  {journal} {Physical Review Letters}\
  }\textbf {\bibinfo {volume} {114}},\ \bibinfo {pages} {096403} (\bibinfo
  {year} {2015})}\BibitemShut {NoStop}%
\bibitem [{\citenamefont {Martins}\ \emph {et~al.}(2011)\citenamefont
  {Martins}, \citenamefont {Aichhorn}, \citenamefont {Vaugier},\ and\
  \citenamefont {Biermann}}]{martins2011reduced}%
  \BibitemOpen
  \bibfield  {author} {\bibinfo {author} {\bibfnamefont {C}~\bibnamefont
  {Martins}}, \bibinfo {author} {\bibfnamefont {M}~\bibnamefont {Aichhorn}},
  \bibinfo {author} {\bibfnamefont {L}~\bibnamefont {Vaugier}}, \ and\ \bibinfo
  {author} {\bibfnamefont {S}~\bibnamefont {Biermann}},\ }\bibfield  {title}
  {\enquote {\bibinfo {title} {Reduced effective spin-orbital degeneracy and
  spin-orbital ordering in paramagnetic transition-metal oxides:
  {Sr$_2$IrO$_4$} versus {Sr$_2$RhO$_4$}},}\ }\href@noop {} {\bibfield
  {journal} {\bibinfo  {journal} {Physical Review Letters}\ }\textbf {\bibinfo
  {volume} {107}},\ \bibinfo {pages} {266404} (\bibinfo {year}
  {2011})}\BibitemShut {NoStop}%
\bibitem [{\citenamefont {Luo}\ \emph {et~al.}(2013)\citenamefont {Luo},
  \citenamefont {Cao}, \citenamefont {Si}, \citenamefont {Li}, \citenamefont
  {Bao}, \citenamefont {Guo}, \citenamefont {Yang}, \citenamefont {Shen},
  \citenamefont {Feng}, \citenamefont {Dai} \emph {et~al.}}]{luo2013li}%
  \BibitemOpen
  \bibfield  {author} {\bibinfo {author} {\bibfnamefont {Y}~\bibnamefont
  {Luo}}, \bibinfo {author} {\bibfnamefont {C}~\bibnamefont {Cao}}, \bibinfo
  {author} {\bibfnamefont {B}~\bibnamefont {Si}}, \bibinfo {author}
  {\bibfnamefont {Y}~\bibnamefont {Li}}, \bibinfo {author} {\bibfnamefont
  {J}~\bibnamefont {Bao}}, \bibinfo {author} {\bibfnamefont {H}~\bibnamefont
  {Guo}}, \bibinfo {author} {\bibfnamefont {X}~\bibnamefont {Yang}}, \bibinfo
  {author} {\bibfnamefont {C}~\bibnamefont {Shen}}, \bibinfo {author}
  {\bibfnamefont {C}~\bibnamefont {Feng}}, \bibinfo {author} {\bibfnamefont
  {J}~\bibnamefont {Dai}},  \emph {et~al.},\ }\bibfield  {title} {\enquote
  {\bibinfo {title} {{Li$_2$RhO$_3$}: A spin-glassy relativistic {Mott}
  insulator},}\ }\href@noop {} {\bibfield  {journal} {\bibinfo  {journal}
  {Physical Review B}\ }\textbf {\bibinfo {volume} {87}},\ \bibinfo {pages}
  {161121} (\bibinfo {year} {2013})}\BibitemShut {NoStop}%
\bibitem [{\citenamefont {Wu}\ \emph {et~al.}(2009)\citenamefont {Wu},
  \citenamefont {McCollam}, \citenamefont {Swainson}, \citenamefont
  {Rancourt},\ and\ \citenamefont {Julian}}]{wu2009novel}%
  \BibitemOpen
  \bibfield  {author} {\bibinfo {author} {\bibfnamefont {W}~\bibnamefont {Wu}},
  \bibinfo {author} {\bibfnamefont {A}~\bibnamefont {McCollam}}, \bibinfo
  {author} {\bibfnamefont {I}~\bibnamefont {Swainson}}, \bibinfo {author}
  {\bibfnamefont {DG}~\bibnamefont {Rancourt}}, \ and\ \bibinfo {author}
  {\bibfnamefont {SR}~\bibnamefont {Julian}},\ }\bibfield  {title} {\enquote
  {\bibinfo {title} {A novel non--{Fermi}-liquid state in the iron-pnictide
  {FeCrAs}},}\ }\href@noop {} {\bibfield  {journal} {\bibinfo  {journal}
  {Europhysics Letters}\ }\textbf {\bibinfo {volume} {85}},\ \bibinfo {pages}
  {17009} (\bibinfo {year} {2009})}\BibitemShut {NoStop}%
\bibitem [{\citenamefont {Akrap}\ \emph {et~al.}(2014)\citenamefont {Akrap},
  \citenamefont {Dai}, \citenamefont {Wu}, \citenamefont {Julian},\ and\
  \citenamefont {Homes}}]{akrap2014optical}%
  \BibitemOpen
  \bibfield  {author} {\bibinfo {author} {\bibfnamefont {A}~\bibnamefont
  {Akrap}}, \bibinfo {author} {\bibfnamefont {YM}~\bibnamefont {Dai}}, \bibinfo
  {author} {\bibfnamefont {W}~\bibnamefont {Wu}}, \bibinfo {author}
  {\bibfnamefont {SR}~\bibnamefont {Julian}}, \ and\ \bibinfo {author}
  {\bibfnamefont {CC}~\bibnamefont {Homes}},\ }\bibfield  {title} {\enquote
  {\bibinfo {title} {Optical properties and electronic structure of the
  nonmetallic metal {FeCrAs}},}\ }\href@noop {} {\bibfield  {journal} {\bibinfo
   {journal} {Physical Review B}\ }\textbf {\bibinfo {volume} {89}},\ \bibinfo
  {pages} {125115} (\bibinfo {year} {2014})}\BibitemShut {NoStop}%
\bibitem [{\citenamefont {Muller}\ and\ \citenamefont
  {Roy}(1968)}]{muller1968formation}%
  \BibitemOpen
  \bibfield  {author} {\bibinfo {author} {\bibfnamefont {O}~\bibnamefont
  {Muller}}\ and\ \bibinfo {author} {\bibfnamefont {R}~\bibnamefont {Roy}},\
  }\bibfield  {title} {\enquote {\bibinfo {title} {Formation and stability of
  the platinum and rhodium oxides at high oxygen pressures and the structures
  of {Pt$_3$O$_4$}, {$\beta$-PtO$_2$} and {RhO$_2$}},}\ }\href@noop {}
  {\bibfield  {journal} {\bibinfo  {journal} {Journal of the Less Common
  Metals}\ }\textbf {\bibinfo {volume} {16}},\ \bibinfo {pages} {129--146}
  (\bibinfo {year} {1968})}\BibitemShut {NoStop}%
\bibitem [{\citenamefont {Shannon}(1968)}]{shannon1968synthesis}%
  \BibitemOpen
  \bibfield  {author} {\bibinfo {author} {\bibfnamefont {RD}~\bibnamefont
  {Shannon}},\ }\bibfield  {title} {\enquote {\bibinfo {title} {Synthesis and
  properties of two new members of the rutile family {RhO$_2$} and
  {PtO$_2$}},}\ }\href@noop {} {\bibfield  {journal} {\bibinfo  {journal}
  {Solid State Communications}\ }\textbf {\bibinfo {volume} {6}},\ \bibinfo
  {pages} {139--143} (\bibinfo {year} {1968})}\BibitemShut {NoStop}%
\bibitem [{\citenamefont {Bain}\ and\ \citenamefont
  {Berry}(2008)}]{bain2008diamagnetic}%
  \BibitemOpen
  \bibfield  {author} {\bibinfo {author} {\bibfnamefont {Gordon~A}\
  \bibnamefont {Bain}}\ and\ \bibinfo {author} {\bibfnamefont {John~F}\
  \bibnamefont {Berry}},\ }\bibfield  {title} {\enquote {\bibinfo {title}
  {Diamagnetic corrections and {Pascal's} constants},}\ }\href@noop {}
  {\bibfield  {journal} {\bibinfo  {journal} {Journal of Chemical Education}\
  }\textbf {\bibinfo {volume} {85}},\ \bibinfo {pages} {532} (\bibinfo {year}
  {2008})}\BibitemShut {NoStop}%
\bibitem [{\citenamefont {Yamaura}\ and\ \citenamefont
  {Takayama-Muromachi}(2001)}]{yamaura2001enhanced}%
  \BibitemOpen
  \bibfield  {author} {\bibinfo {author} {\bibfnamefont {K}~\bibnamefont
  {Yamaura}}\ and\ \bibinfo {author} {\bibfnamefont {E}~\bibnamefont
  {Takayama-Muromachi}},\ }\bibfield  {title} {\enquote {\bibinfo {title}
  {Enhanced paramagnetism of the {$4d$} itinerant electrons in the rhodium
  oxide perovskite {SrRhO$_3$}},}\ }\href@noop {} {\bibfield  {journal}
  {\bibinfo  {journal} {Physical Review B}\ }\textbf {\bibinfo {volume} {64}},\
  \bibinfo {pages} {224424} (\bibinfo {year} {2001})}\BibitemShut {NoStop}%
\bibitem [{\citenamefont {Singh}(2003)}]{singh2003prospects}%
  \BibitemOpen
  \bibfield  {author} {\bibinfo {author} {\bibfnamefont {David~J}\ \bibnamefont
  {Singh}},\ }\bibfield  {title} {\enquote {\bibinfo {title} {Prospects for
  quantum criticality in perovskite {SrRhO$_3$}},}\ }\href@noop {} {\bibfield
  {journal} {\bibinfo  {journal} {Physical Review B}\ }\textbf {\bibinfo
  {volume} {67}},\ \bibinfo {pages} {054507} (\bibinfo {year}
  {2003})}\BibitemShut {NoStop}%
\bibitem [{\citenamefont {Saha}\ \emph {et~al.}(2008)\citenamefont {Saha},
  \citenamefont {Singh}, \citenamefont {Dkhil}, \citenamefont {Dhar},
  \citenamefont {Suryanarayanan}, \citenamefont {Dhalenne}, \citenamefont
  {Revcolevschi},\ and\ \citenamefont {Sood}}]{saha2008temperature}%
  \BibitemOpen
  \bibfield  {author} {\bibinfo {author} {\bibfnamefont {S}~\bibnamefont
  {Saha}}, \bibinfo {author} {\bibfnamefont {S}~\bibnamefont {Singh}}, \bibinfo
  {author} {\bibfnamefont {B}~\bibnamefont {Dkhil}}, \bibinfo {author}
  {\bibfnamefont {S}~\bibnamefont {Dhar}}, \bibinfo {author} {\bibfnamefont
  {R}~\bibnamefont {Suryanarayanan}}, \bibinfo {author} {\bibfnamefont
  {G}~\bibnamefont {Dhalenne}}, \bibinfo {author} {\bibfnamefont
  {A}~\bibnamefont {Revcolevschi}}, \ and\ \bibinfo {author} {\bibfnamefont
  {AK}~\bibnamefont {Sood}},\ }\bibfield  {title} {\enquote {\bibinfo {title}
  {Temperature-dependent {Raman} and x-ray studies of the spin-ice pyrochlore
  {Dy$_2$Ti$_2$O$_7$} and nonmagnetic pyrochlore {Lu$_2$Ti$_2$O$_7$}},}\
  }\href@noop {} {\bibfield  {journal} {\bibinfo  {journal} {Physical Review
  B}\ }\textbf {\bibinfo {volume} {78}},\ \bibinfo {pages} {214102} (\bibinfo
  {year} {2008})}\BibitemShut {NoStop}%
\bibitem [{\citenamefont {Maczka}\ \emph {et~al.}(2008)\citenamefont {Maczka},
  \citenamefont {Hanuza}, \citenamefont {Hermanowicz}, \citenamefont {Fuentes},
  \citenamefont {Matsuhira},\ and\ \citenamefont
  {Hiroi}}]{maczka2008temperature}%
  \BibitemOpen
  \bibfield  {author} {\bibinfo {author} {\bibfnamefont {M}~\bibnamefont
  {Maczka}}, \bibinfo {author} {\bibfnamefont {J}~\bibnamefont {Hanuza}},
  \bibinfo {author} {\bibfnamefont {K}~\bibnamefont {Hermanowicz}}, \bibinfo
  {author} {\bibfnamefont {AF}~\bibnamefont {Fuentes}}, \bibinfo {author}
  {\bibfnamefont {K}~\bibnamefont {Matsuhira}}, \ and\ \bibinfo {author}
  {\bibfnamefont {Z}~\bibnamefont {Hiroi}},\ }\bibfield  {title} {\enquote
  {\bibinfo {title} {Temperature-dependent {Raman} scattering studies of the
  geometrically frustrated pyrochlores {Dy$_2$Ti$_2$O$_7$}, {Gd$_2$Ti$_2$O$_7$}
  and {Er$_2$Ti$_2$O$_7$}},}\ }\href@noop {} {\bibfield  {journal} {\bibinfo
  {journal} {Journal of Raman Spectroscopy}\ }\textbf {\bibinfo {volume}
  {39}},\ \bibinfo {pages} {537--544} (\bibinfo {year} {2008})}\BibitemShut
  {NoStop}%
\bibitem [{\citenamefont {Maeno}\ \emph {et~al.}(1997)\citenamefont {Maeno},
  \citenamefont {Yoshida}, \citenamefont {Hashimoto}, \citenamefont
  {Nishizaki}, \citenamefont {Ikeda}, \citenamefont {Nohara}, \citenamefont
  {Fujita}, \citenamefont {Mackenzie}, \citenamefont {Hussey}, \citenamefont
  {Bednorz},\ and\ \citenamefont {Lichtenberg}}]{maeno1997two}%
  \BibitemOpen
  \bibfield  {author} {\bibinfo {author} {\bibfnamefont {Y}~\bibnamefont
  {Maeno}}, \bibinfo {author} {\bibfnamefont {K}~\bibnamefont {Yoshida}},
  \bibinfo {author} {\bibfnamefont {H}~\bibnamefont {Hashimoto}}, \bibinfo
  {author} {\bibfnamefont {S}~\bibnamefont {Nishizaki}}, \bibinfo {author}
  {\bibfnamefont {S}~\bibnamefont {Ikeda}}, \bibinfo {author} {\bibfnamefont
  {M}~\bibnamefont {Nohara}}, \bibinfo {author} {\bibfnamefont {T}~\bibnamefont
  {Fujita}}, \bibinfo {author} {\bibfnamefont {AP}~\bibnamefont {Mackenzie}},
  \bibinfo {author} {\bibfnamefont {NE}~\bibnamefont {Hussey}}, \bibinfo
  {author} {\bibfnamefont {JG}~\bibnamefont {Bednorz}}, \ and\ \bibinfo
  {author} {\bibfnamefont {F}~\bibnamefont {Lichtenberg}},\ }\bibfield  {title}
  {\enquote {\bibinfo {title} {Two-dimensional {Fermi} liquid behavior of the
  superconductor {Sr$_2$RuO$_4$}},}\ }\href@noop {} {\bibfield  {journal}
  {\bibinfo  {journal} {Journal of the Physical Society of Japan}\ }\textbf
  {\bibinfo {volume} {66}},\ \bibinfo {pages} {1405--1408} (\bibinfo {year}
  {1997})}\BibitemShut {NoStop}%
\bibitem [{\citenamefont {Park}\ \emph {et~al.}(2016)\citenamefont {Park},
  \citenamefont {Tan}, \citenamefont {Adroja}, \citenamefont {Daoud-Aladine},
  \citenamefont {Choi}, \citenamefont {Cho}, \citenamefont {Lee}, \citenamefont
  {Kim}, \citenamefont {Sim}, \citenamefont {Morioka}, \citenamefont {Nojiri},
  \citenamefont {Krishnamurthy}, \citenamefont {Manuel}, \citenamefont {Lees},
  \citenamefont {Streltsov}, \citenamefont {Khomskii},\ and\ \citenamefont
  {Park}}]{park2016robust}%
  \BibitemOpen
  \bibfield  {author} {\bibinfo {author} {\bibfnamefont {J}~\bibnamefont
  {Park}}, \bibinfo {author} {\bibfnamefont {T-Y}\ \bibnamefont {Tan}},
  \bibinfo {author} {\bibfnamefont {DT}~\bibnamefont {Adroja}}, \bibinfo
  {author} {\bibfnamefont {A}~\bibnamefont {Daoud-Aladine}}, \bibinfo {author}
  {\bibfnamefont {S}~\bibnamefont {Choi}}, \bibinfo {author} {\bibfnamefont
  {D-Y}\ \bibnamefont {Cho}}, \bibinfo {author} {\bibfnamefont {S-H}\
  \bibnamefont {Lee}}, \bibinfo {author} {\bibfnamefont {J}~\bibnamefont
  {Kim}}, \bibinfo {author} {\bibfnamefont {H}~\bibnamefont {Sim}}, \bibinfo
  {author} {\bibfnamefont {T}~\bibnamefont {Morioka}}, \bibinfo {author}
  {\bibfnamefont {H}~\bibnamefont {Nojiri}}, \bibinfo {author} {\bibfnamefont
  {VV}~\bibnamefont {Krishnamurthy}}, \bibinfo {author} {\bibfnamefont
  {P}~\bibnamefont {Manuel}}, \bibinfo {author} {\bibfnamefont
  {MR}~\bibnamefont {Lees}}, \bibinfo {author} {\bibfnamefont {SV}~\bibnamefont
  {Streltsov}}, \bibinfo {author} {\bibfnamefont {DI}~\bibnamefont {Khomskii}},
  \ and\ \bibinfo {author} {\bibfnamefont {J-G}\ \bibnamefont {Park}},\
  }\bibfield  {title} {\enquote {\bibinfo {title} {Robust singlet dimers with
  fragile ordering in two-dimensional honeycomb lattice of {Li$_2$RuO$_3$}},}\
  }\href@noop {} {\bibfield  {journal} {\bibinfo  {journal} {Scientific
  Reports}\ }\textbf {\bibinfo {volume} {6}},\ \bibinfo {pages} {25238}
  (\bibinfo {year} {2016})}\BibitemShut {NoStop}%
\bibitem [{\citenamefont {Perry}\ \emph {et~al.}(2006)\citenamefont {Perry},
  \citenamefont {Baumberger}, \citenamefont {Balicas}, \citenamefont
  {Kikugawa}, \citenamefont {Ingle}, \citenamefont {Rost}, \citenamefont
  {Mercure}, \citenamefont {Maeno}, \citenamefont {Shen},\ and\ \citenamefont
  {Mackenzie}}]{perry2006sr2rho4}%
  \BibitemOpen
  \bibfield  {author} {\bibinfo {author} {\bibfnamefont {RS}~\bibnamefont
  {Perry}}, \bibinfo {author} {\bibfnamefont {F}~\bibnamefont {Baumberger}},
  \bibinfo {author} {\bibfnamefont {L}~\bibnamefont {Balicas}}, \bibinfo
  {author} {\bibfnamefont {N}~\bibnamefont {Kikugawa}}, \bibinfo {author}
  {\bibfnamefont {NJC}\ \bibnamefont {Ingle}}, \bibinfo {author} {\bibfnamefont
  {A}~\bibnamefont {Rost}}, \bibinfo {author} {\bibfnamefont {JF}~\bibnamefont
  {Mercure}}, \bibinfo {author} {\bibfnamefont {Y}~\bibnamefont {Maeno}},
  \bibinfo {author} {\bibfnamefont {ZX}~\bibnamefont {Shen}}, \ and\ \bibinfo
  {author} {\bibfnamefont {AP}~\bibnamefont {Mackenzie}},\ }\bibfield  {title}
  {\enquote {\bibinfo {title} {{Sr$_2$RhO$_4$}: A new, clean correlated
  electron metal},}\ }\href@noop {} {\bibfield  {journal} {\bibinfo  {journal}
  {New Journal of Physics}\ }\textbf {\bibinfo {volume} {8}},\ \bibinfo {pages}
  {175} (\bibinfo {year} {2006})}\BibitemShut {NoStop}%
\bibitem [{\citenamefont {Stewart}(1984)}]{stewart1984heavy}%
  \BibitemOpen
  \bibfield  {author} {\bibinfo {author} {\bibfnamefont {GR}~\bibnamefont
  {Stewart}},\ }\bibfield  {title} {\enquote {\bibinfo {title} {Heavy-fermion
  systems},}\ }\href@noop {} {\bibfield  {journal} {\bibinfo  {journal}
  {Reviews of Modern Physics}\ }\textbf {\bibinfo {volume} {56}},\ \bibinfo
  {pages} {755} (\bibinfo {year} {1984})}\BibitemShut {NoStop}%
\bibitem [{\citenamefont {Shiga}\ \emph {et~al.}(1993)\citenamefont {Shiga},
  \citenamefont {Fujisawa},\ and\ \citenamefont {Wada}}]{shiga1993spin}%
  \BibitemOpen
  \bibfield  {author} {\bibinfo {author} {\bibfnamefont {Masayuki}\
  \bibnamefont {Shiga}}, \bibinfo {author} {\bibfnamefont {Koji}\ \bibnamefont
  {Fujisawa}}, \ and\ \bibinfo {author} {\bibfnamefont {Hirofumi}\ \bibnamefont
  {Wada}},\ }\bibfield  {title} {\enquote {\bibinfo {title} {Spin liquid
  behavior of highly frustrated {Y(Sc)Mn$_2$} and effects of nonmagnetic
  impurity},}\ }\href@noop {} {\bibfield  {journal} {\bibinfo  {journal}
  {Journal of the Physical Society of Japan}\ }\textbf {\bibinfo {volume}
  {62}},\ \bibinfo {pages} {1329--1336} (\bibinfo {year} {1993})}\BibitemShut
  {NoStop}%
\bibitem [{\citenamefont {Kondo}\ \emph {et~al.}(1997)\citenamefont {Kondo},
  \citenamefont {Johnston}, \citenamefont {Swenson}, \citenamefont {Borsa},
  \citenamefont {Mahajan}, \citenamefont {Miller}, \citenamefont {Gu},
  \citenamefont {Goldman}, \citenamefont {Maple}, \citenamefont {Gajewski}
  \emph {et~al.}}]{kondo1997liv}%
  \BibitemOpen
  \bibfield  {author} {\bibinfo {author} {\bibfnamefont {Shinichiro}\
  \bibnamefont {Kondo}}, \bibinfo {author} {\bibfnamefont {DC}~\bibnamefont
  {Johnston}}, \bibinfo {author} {\bibfnamefont {CA}~\bibnamefont {Swenson}},
  \bibinfo {author} {\bibfnamefont {F}~\bibnamefont {Borsa}}, \bibinfo {author}
  {\bibfnamefont {AV}~\bibnamefont {Mahajan}}, \bibinfo {author} {\bibfnamefont
  {LL}~\bibnamefont {Miller}}, \bibinfo {author} {\bibfnamefont
  {T}~\bibnamefont {Gu}}, \bibinfo {author} {\bibfnamefont {AI}~\bibnamefont
  {Goldman}}, \bibinfo {author} {\bibfnamefont {MB}~\bibnamefont {Maple}},
  \bibinfo {author} {\bibfnamefont {DA}~\bibnamefont {Gajewski}},  \emph
  {et~al.},\ }\bibfield  {title} {\enquote {\bibinfo {title} {{LiV$_2$O$_4$}: A
  heavy fermion transition metal oxide},}\ }\href@noop {} {\bibfield  {journal}
  {\bibinfo  {journal} {Physical Review Letters}\ }\textbf {\bibinfo {volume}
  {78}},\ \bibinfo {pages} {3729} (\bibinfo {year} {1997})}\BibitemShut
  {NoStop}%
\bibitem [{\citenamefont {Lacroix}(2001)}]{lacroix2001heavy}%
  \BibitemOpen
  \bibfield  {author} {\bibinfo {author} {\bibfnamefont {C}~\bibnamefont
  {Lacroix}},\ }\bibfield  {title} {\enquote {\bibinfo {title} {Heavy-fermion
  behavior of itinerant frustrated systems: {$\beta$-Mn, Y(Sc)Mn$_2$, and
  LiV$_2$O$_4$}},}\ }\href@noop {} {\bibfield  {journal} {\bibinfo  {journal}
  {Canadian journal of physics}\ }\textbf {\bibinfo {volume} {79}},\ \bibinfo
  {pages} {1469--1473} (\bibinfo {year} {2001})}\BibitemShut {NoStop}%
\bibitem [{\citenamefont {Wada}\ \emph {et~al.}(1987)\citenamefont {Wada},
  \citenamefont {Nakamura}, \citenamefont {Fukami}, \citenamefont {Yoshimura},
  \citenamefont {Shiga},\ and\ \citenamefont {Nakamura}}]{wada1987spin}%
  \BibitemOpen
  \bibfield  {author} {\bibinfo {author} {\bibfnamefont {H}~\bibnamefont
  {Wada}}, \bibinfo {author} {\bibfnamefont {H}~\bibnamefont {Nakamura}},
  \bibinfo {author} {\bibfnamefont {E}~\bibnamefont {Fukami}}, \bibinfo
  {author} {\bibfnamefont {K}~\bibnamefont {Yoshimura}}, \bibinfo {author}
  {\bibfnamefont {M}~\bibnamefont {Shiga}}, \ and\ \bibinfo {author}
  {\bibfnamefont {Y}~\bibnamefont {Nakamura}},\ }\bibfield  {title} {\enquote
  {\bibinfo {title} {Spin fluctuations of {Y(Mn$_{1-x}$Al$_x$)$_2$} and
  {Y$_{1-x}$Sc$_x$Mn$_2$}},}\ }\href@noop {} {\bibfield  {journal} {\bibinfo
  {journal} {Journal of magnetism and magnetic materials}\ }\textbf {\bibinfo
  {volume} {70}},\ \bibinfo {pages} {17--19} (\bibinfo {year}
  {1987})}\BibitemShut {NoStop}%
\bibitem [{\citenamefont {Palstra}\ \emph {et~al.}(1985)\citenamefont
  {Palstra}, \citenamefont {Menovsky}, \citenamefont {Van~den Berg},
  \citenamefont {Dirkmaat}, \citenamefont {Kes}, \citenamefont {Nieuwenhuys},\
  and\ \citenamefont {Mydosh}}]{palstra1985superconducting}%
  \BibitemOpen
  \bibfield  {author} {\bibinfo {author} {\bibfnamefont {TTM}\ \bibnamefont
  {Palstra}}, \bibinfo {author} {\bibfnamefont {AA}~\bibnamefont {Menovsky}},
  \bibinfo {author} {\bibfnamefont {J}~\bibnamefont {Van~den Berg}}, \bibinfo
  {author} {\bibfnamefont {AJ}~\bibnamefont {Dirkmaat}}, \bibinfo {author}
  {\bibfnamefont {PH}~\bibnamefont {Kes}}, \bibinfo {author} {\bibfnamefont
  {GJ}~\bibnamefont {Nieuwenhuys}}, \ and\ \bibinfo {author} {\bibfnamefont
  {JA}~\bibnamefont {Mydosh}},\ }\bibfield  {title} {\enquote {\bibinfo {title}
  {Superconducting and magnetic transitions in the heavy-fermion system
  {URu$_2$Si$_2$}},}\ }\href@noop {} {\bibfield  {journal} {\bibinfo  {journal}
  {Physical Review Letters}\ }\textbf {\bibinfo {volume} {55}},\ \bibinfo
  {pages} {2727} (\bibinfo {year} {1985})}\BibitemShut {NoStop}%
\bibitem [{\citenamefont {Frings}\ \emph {et~al.}(1983)\citenamefont {Frings},
  \citenamefont {Franse}, \citenamefont {De~Boer},\ and\ \citenamefont
  {Menovsky}}]{frings1983magnetic}%
  \BibitemOpen
  \bibfield  {author} {\bibinfo {author} {\bibfnamefont {PH}~\bibnamefont
  {Frings}}, \bibinfo {author} {\bibfnamefont {JJM}\ \bibnamefont {Franse}},
  \bibinfo {author} {\bibfnamefont {FR}~\bibnamefont {De~Boer}}, \ and\
  \bibinfo {author} {\bibfnamefont {A}~\bibnamefont {Menovsky}},\ }\bibfield
  {title} {\enquote {\bibinfo {title} {Magnetic properties of {U$_x$Pt$_y$}
  compounds},}\ }\href@noop {} {\bibfield  {journal} {\bibinfo  {journal}
  {Journal of magnetism and magnetic materials}\ }\textbf {\bibinfo {volume}
  {31}},\ \bibinfo {pages} {240--242} (\bibinfo {year} {1983})}\BibitemShut
  {NoStop}%
\bibitem [{\citenamefont {Zhou}\ \emph {et~al.}(2003)\citenamefont {Zhou},
  \citenamefont {Goodenough},\ and\ \citenamefont
  {Dabrowski}}]{zhou2003transition}%
  \BibitemOpen
  \bibfield  {author} {\bibinfo {author} {\bibfnamefont {J-S}\ \bibnamefont
  {Zhou}}, \bibinfo {author} {\bibfnamefont {JB}~\bibnamefont {Goodenough}}, \
  and\ \bibinfo {author} {\bibfnamefont {B}~\bibnamefont {Dabrowski}},\
  }\bibfield  {title} {\enquote {\bibinfo {title} {Transition from
  {Curie-Weiss} to enhanced {Pauli} paramagnetism in {$R$NiO$_3$ ($R=$ La, Pr,
  … Gd)}},}\ }\href@noop {} {\bibfield  {journal} {\bibinfo  {journal}
  {Physical Review B}\ }\textbf {\bibinfo {volume} {67}},\ \bibinfo {pages}
  {020404} (\bibinfo {year} {2003})}\BibitemShut {NoStop}%
\bibitem [{\citenamefont {Yamaura}\ \emph {et~al.}(2002)\citenamefont
  {Yamaura}, \citenamefont {Huang}, \citenamefont {Young}, \citenamefont
  {Noguchi},\ and\ \citenamefont {Takayama-Muromachi}}]{yamaura2002crystal}%
  \BibitemOpen
  \bibfield  {author} {\bibinfo {author} {\bibfnamefont {K}~\bibnamefont
  {Yamaura}}, \bibinfo {author} {\bibfnamefont {Q}~\bibnamefont {Huang}},
  \bibinfo {author} {\bibfnamefont {DP}~\bibnamefont {Young}}, \bibinfo
  {author} {\bibfnamefont {Y}~\bibnamefont {Noguchi}}, \ and\ \bibinfo {author}
  {\bibfnamefont {E}~\bibnamefont {Takayama-Muromachi}},\ }\bibfield  {title}
  {\enquote {\bibinfo {title} {Crystal structure and electronic and magnetic
  properties of the bilayered rhodium oxide {Sr$_3$Rh$_2$O$_7$}},}\ }\href@noop
  {} {\bibfield  {journal} {\bibinfo  {journal} {Physical Review B}\ }\textbf
  {\bibinfo {volume} {66}},\ \bibinfo {pages} {134431} (\bibinfo {year}
  {2002})}\BibitemShut {NoStop}%
\bibitem [{\citenamefont {Yamaura}\ \emph {et~al.}(2004)\citenamefont
  {Yamaura}, \citenamefont {Huang}, \citenamefont {Young},\ and\ \citenamefont
  {Takayama-Muromachi}}]{yamaura2004crystal}%
  \BibitemOpen
  \bibfield  {author} {\bibinfo {author} {\bibfnamefont {K}~\bibnamefont
  {Yamaura}}, \bibinfo {author} {\bibfnamefont {Q}~\bibnamefont {Huang}},
  \bibinfo {author} {\bibfnamefont {DP}~\bibnamefont {Young}}, \ and\ \bibinfo
  {author} {\bibfnamefont {E}~\bibnamefont {Takayama-Muromachi}},\ }\bibfield
  {title} {\enquote {\bibinfo {title} {Crystal structure and magnetic
  properties of the trilayered perovskite {Sr$_4$Rh$_3$O$_{10}$}: A new member
  of the strontium rhodate family},}\ }\href@noop {} {\bibfield  {journal}
  {\bibinfo  {journal} {Chemistry of Materials}\ }\textbf {\bibinfo {volume}
  {16}},\ \bibinfo {pages} {3424--3430} (\bibinfo {year} {2004})}\BibitemShut
  {NoStop}%
\bibitem [{\citenamefont {Yamaura}\ \emph {et~al.}(2005)\citenamefont
  {Yamaura}, \citenamefont {Huang}, \citenamefont {Moldovan}, \citenamefont
  {Young}, \citenamefont {Sato}, \citenamefont {Baba}, \citenamefont {Nagai},
  \citenamefont {Matsui},\ and\ \citenamefont
  {Takayama-Muromachi}}]{yamaura2005high}%
  \BibitemOpen
  \bibfield  {author} {\bibinfo {author} {\bibfnamefont {K}~\bibnamefont
  {Yamaura}}, \bibinfo {author} {\bibfnamefont {Q}~\bibnamefont {Huang}},
  \bibinfo {author} {\bibfnamefont {M}~\bibnamefont {Moldovan}}, \bibinfo
  {author} {\bibfnamefont {DP}~\bibnamefont {Young}}, \bibinfo {author}
  {\bibfnamefont {A}~\bibnamefont {Sato}}, \bibinfo {author} {\bibfnamefont
  {Y}~\bibnamefont {Baba}}, \bibinfo {author} {\bibfnamefont {T}~\bibnamefont
  {Nagai}}, \bibinfo {author} {\bibfnamefont {Y}~\bibnamefont {Matsui}}, \ and\
  \bibinfo {author} {\bibfnamefont {E}~\bibnamefont {Takayama-Muromachi}},\
  }\bibfield  {title} {\enquote {\bibinfo {title} {High-pressure synthesis,
  crystal structure determination, and a {Ca} substitution study of the
  metallic rhodium oxide {NaRh$_2$O$_4$}},}\ }\href@noop {} {\bibfield
  {journal} {\bibinfo  {journal} {Chemistry of Materials}\ }\textbf {\bibinfo
  {volume} {17}},\ \bibinfo {pages} {359--365} (\bibinfo {year}
  {2005})}\BibitemShut {NoStop}%
\bibitem [{\citenamefont {Yamaura}\ and\ \citenamefont
  {Takayama-Muromachi}(2006)}]{yamaura2006high}%
  \BibitemOpen
  \bibfield  {author} {\bibinfo {author} {\bibfnamefont {K}~\bibnamefont
  {Yamaura}}\ and\ \bibinfo {author} {\bibfnamefont {E}~\bibnamefont
  {Takayama-Muromachi}},\ }\bibfield  {title} {\enquote {\bibinfo {title}
  {High-pressure synthesis of the perovskite rhodate {CaRhO$_3$}},}\
  }\href@noop {} {\bibfield  {journal} {\bibinfo  {journal} {Physica C}\
  }\textbf {\bibinfo {volume} {445}},\ \bibinfo {pages} {54--56} (\bibinfo
  {year} {2006})}\BibitemShut {NoStop}%
\bibitem [{\citenamefont {Rau}\ and\ \citenamefont
  {Kee}(2011)}]{rau2011hidden}%
  \BibitemOpen
  \bibfield  {author} {\bibinfo {author} {\bibfnamefont {JG}~\bibnamefont
  {Rau}}\ and\ \bibinfo {author} {\bibfnamefont {H-Y}\ \bibnamefont {Kee}},\
  }\bibfield  {title} {\enquote {\bibinfo {title} {Hidden spin liquid in an
  antiferromagnet: Applications to {FeCrAs}},}\ }\href@noop {} {\bibfield
  {journal} {\bibinfo  {journal} {Physical Review B}\ }\textbf {\bibinfo
  {volume} {84}},\ \bibinfo {pages} {104448} (\bibinfo {year}
  {2011})}\BibitemShut {NoStop}%
\bibitem [{\citenamefont {Nevidomskyy}\ and\ \citenamefont
  {Coleman}(2009)}]{nevidomskyy2009kondo}%
  \BibitemOpen
  \bibfield  {author} {\bibinfo {author} {\bibfnamefont {Andriy~H}\
  \bibnamefont {Nevidomskyy}}\ and\ \bibinfo {author} {\bibfnamefont {Piers}\
  \bibnamefont {Coleman}},\ }\bibfield  {title} {\enquote {\bibinfo {title}
  {Kondo resonance narrowing in $d$- and $f$-electron systems},}\ }\href@noop
  {} {\bibfield  {journal} {\bibinfo  {journal} {Physical Review Letters}\
  }\textbf {\bibinfo {volume} {103}},\ \bibinfo {pages} {147205} (\bibinfo
  {year} {2009})}\BibitemShut {NoStop}%
\bibitem [{\citenamefont {Georges}\ \emph {et~al.}(2013)\citenamefont
  {Georges}, \citenamefont {Medici},\ and\ \citenamefont
  {Mravlje}}]{georges2013strong}%
  \BibitemOpen
  \bibfield  {author} {\bibinfo {author} {\bibfnamefont {A}~\bibnamefont
  {Georges}}, \bibinfo {author} {\bibfnamefont {L~de'}\ \bibnamefont {Medici}},
  \ and\ \bibinfo {author} {\bibfnamefont {J}~\bibnamefont {Mravlje}},\
  }\bibfield  {title} {\enquote {\bibinfo {title} {Strong correlations from
  {Hund’s} coupling},}\ }\href@noop {} {\bibfield  {journal} {\bibinfo
  {journal} {Annu. Rev. Condens. Matter Phys.}\ }\textbf {\bibinfo {volume}
  {4}},\ \bibinfo {pages} {137--178} (\bibinfo {year} {2013})}\BibitemShut
  {NoStop}%
\bibitem [{\citenamefont {Plumb}\ \emph {et~al.}(2018)\citenamefont {Plumb},
  \citenamefont {Stock}, \citenamefont {Rodriguez-Rivera}, \citenamefont
  {Castellan}, \citenamefont {Taylor}, \citenamefont {Lau}, \citenamefont {Wu},
  \citenamefont {Julian},\ and\ \citenamefont {Kim}}]{plumb2018mean}%
  \BibitemOpen
  \bibfield  {author} {\bibinfo {author} {\bibfnamefont {KW}~\bibnamefont
  {Plumb}}, \bibinfo {author} {\bibfnamefont {C}~\bibnamefont {Stock}},
  \bibinfo {author} {\bibfnamefont {JA}~\bibnamefont {Rodriguez-Rivera}},
  \bibinfo {author} {\bibfnamefont {J-P}\ \bibnamefont {Castellan}}, \bibinfo
  {author} {\bibfnamefont {JW}~\bibnamefont {Taylor}}, \bibinfo {author}
  {\bibfnamefont {B}~\bibnamefont {Lau}}, \bibinfo {author} {\bibfnamefont
  {W}~\bibnamefont {Wu}}, \bibinfo {author} {\bibfnamefont {SR}~\bibnamefont
  {Julian}}, \ and\ \bibinfo {author} {\bibfnamefont {Young-June}\ \bibnamefont
  {Kim}},\ }\bibfield  {title} {\enquote {\bibinfo {title} {From mean-field
  localized magnetism to itinerant spin fluctuations in the “nonmetallic
  metal” {FeCrAs}},}\ }\href@noop {} {\bibfield  {journal} {\bibinfo
  {journal} {Physical Review B}\ }\textbf {\bibinfo {volume} {97}},\ \bibinfo
  {pages} {184431} (\bibinfo {year} {2018})}\BibitemShut {NoStop}%
\bibitem [{\citenamefont
  {Rodr{\'\i}guez-Carvajal}(1993)}]{rodriguez1993recent}%
  \BibitemOpen
  \bibfield  {author} {\bibinfo {author} {\bibfnamefont {Juan}\ \bibnamefont
  {Rodr{\'\i}guez-Carvajal}},\ }\bibfield  {title} {\enquote {\bibinfo {title}
  {Recent advances in magnetic structure determination by neutron powder
  diffraction},}\ }\href@noop {} {\bibfield  {journal} {\bibinfo  {journal}
  {Physica B: Condensed Matter}\ }\textbf {\bibinfo {volume} {192}},\ \bibinfo
  {pages} {55--69} (\bibinfo {year} {1993})}\BibitemShut {NoStop}%
\end{thebibliography}%

\newpage
\begin{table}
\caption{Refined lattice positions and temperature factors for Lu$_2$Rh$_2$O$_7$ where the goodness of fit parameters are R$_p$ = 7.34, R$_{wp}$ = 10.9, R$_{exp}$ = 6.27 and $\chi^2$ = 3.02}
\vspace{1em}
\begin{tabular}{lp{1.7cm}<{\centering}p{1.3cm}<{\centering}p{1.3cm}<{\centering}p{1.5cm}<{\centering}p{1.3cm}<{\centering}}
 \hline
 \hline & $x$ & $y$ & $z$ & $B_{iso}$ & site \\
 \hline
 Lu & 0.5 & 0.5 & 0.5 & 0.52(4) & $16d$ \\
 Rh & 0 & 0 & 0 & 0.28(5) & $16c$ \\
 O1 & 0.375 & 0.375 & 0.375 & 0.4(2) & $8b$ \\
 O2 & 0.3296(5) & 0.125 & 0.125 & 0.4(2) & $48f$ \\
 \hline
 \hline
\end{tabular}
\end{table}

\begin{figure}[]
\includegraphics[width=4.2in]{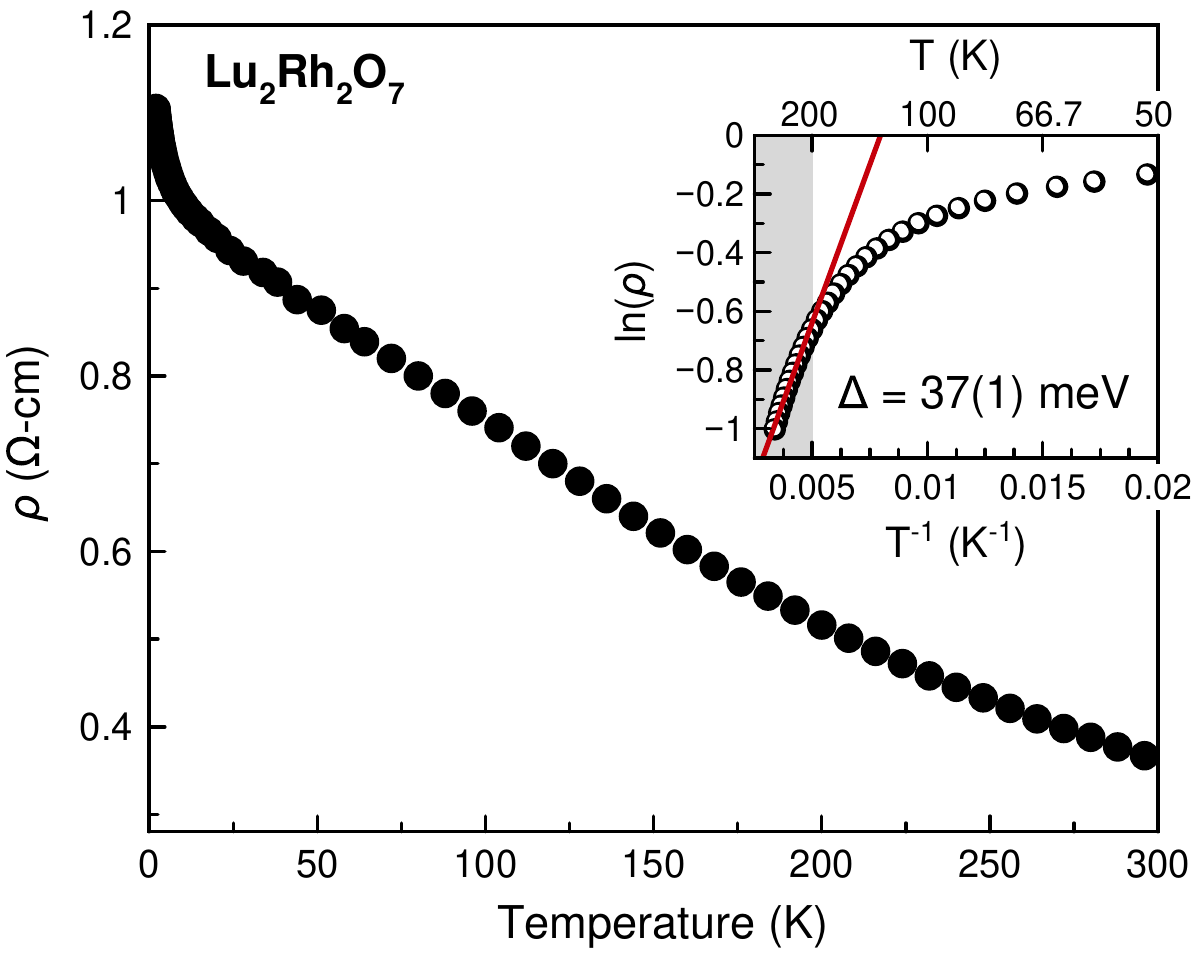} 
\caption{The electrical resistivity of Lu$_2$Rh$_2$O$_7$ is typical of a semiconductor in magnitude but with an abnormal temperature dependence. The inset shows $\ln{(\rho)}$ vs. $T^{-1}$, where the red line indicates a fit to thermally activated Arrhenius behavior yielding a gap of $\Delta = 37(1)$~meV.}
\label{Resistivity}
\end{figure}

\end{document}